\documentclass[preprint,aps,amsmath,amssymb,12pt,showkeys,superscriptaddress]{revtex4} 

\usepackage{hyperref}
\usepackage{bbm}
\usepackage[pdftex]{graphicx}
\usepackage{times}
\usepackage{color}
\usepackage{placeins}
\usepackage{amsmath,amssymb}

\newcommand{\G}{\mathrm{G}}
\newcommand{\q}{\mathrm{q}}
\renewcommand{\v}{\mathrm{v}}

\newcommand{\micron}{{\mu\mathrm{m}}}
\renewcommand{\ol}[1]{\overline{#1}}
\newcommand{\kyaw}{k}
\newcommand{\kpivo}{\ol{k}}

\newcommand{\forcenewpage}{\newpage\FloatBarrier}

\begin{document}

\bibliographystyle{apsrev}

\title{Cell body rocking is a dominant mechanism for flagellar synchronization in a swimming alga}
\author{Veikko Geyer}
\affiliation{Max Planck Institute of Molecular Cell Biology and Genetics}
\author{Frank J\"ulicher}
\affiliation{Max Planck Institute for the Physics of Complex Systems}
\author{Jonathon Howard}
\affiliation{Max Planck Institute of Molecular Cell Biology and Genetics}
\author{Benjamin M Friedrich}
\affiliation{Max Planck Institute for the Physics of Complex Systems}


\begin{abstract} 
The unicellular green algae \textit{Chlamydomonas} swims with two flagella, which can synchronize their beat.
Synchronized beating is required to swim both fast and straight.
A long-standing hypothesis proposes that synchronization of flagella results from hydrodynamic coupling,
but the details are not understood.
Here, we present realistic hydrodynamic computations and high-speed tracking experiments of swimming cells
that show how a perturbation from the synchronized state causes rotational motion of the cell body.
This rotation feeds back on the flagellar dynamics via hydrodynamic friction forces
and rapidly restores the synchronized state in our theory.
We calculate that this `cell body rocking' provides the dominant contribution to synchronization 
in swimming cells,
whereas direct hydrodynamic interactions between the flagella contribute negligibly.
We experimentally confirmed the coupling between flagellar beating and cell body rocking predicted by our theory.
We propose that the interplay of flagellar beating and hydrodynamic forces
governs swimming and synchronization in \textit{Chlamydomonas}.

\vspace{1cm}
This work appeared also in the Proceedings of the National Academy of Science of the U.S.A as:

\href{http://www.pnas.org/content/early/2013/10/17/1300895110.abstract}{Geyer \textit{et al.}, Proc. Natl. Acad. Sci. U.S.A., 
\textbf{110}(45), p.~18058(6), 2013.}

\end{abstract}


\maketitle

Eukaryotic cilia and flagella are long, slender cell appendages that can bend rhythmically
and thus represent a prime example of a biological oscillator \cite{Alberts:cell}.
The flagellar beat is driven by the collective action of dynein molecular motors,
which are distributed along the length of the flagellum.
The beat of flagella, with typical frequencies ranging from $20-60\,\mathrm{Hz}$,
pumps fluids, \textit{e.g.} mucus in mammalian airways \cite{Sanderson:1981},
and propels unicellular micro-swimmers like \textit{Paramecium}, spermatozoa, or algae \cite{Gray:1928}.
The coordinated beating of collections of flagella is important for efficient fluid transport 
\cite{Sanderson:1981,Osterman:2011} and fast swimming \cite{Brennen:1977}.
This coordinated beating represents a striking example for the synchronization of oscillators,
prompting the question of how flagella couple their beat.
Identifying the specific mechanism of synchronization can be difficult as
synchronization may occur even for weak coupling \cite{Pikovsky:synchronization}.
Further, the effect of the coupling is difficult to detect once the synchronized state has been reached.

Hydrodynamic forces were suggested to play a significant role for flagellar synchronization
already in 1951 by G.I.~Taylor \cite{Taylor:1951}.
Since then, direct hydrodynamic interactions between flagella were studied theoretically 
as a possible mechanism for flagellar synchronization \cite{Gueron:1999,Vilfan:2006,Niedermayer:2008,Golestanian:2011}.
Another synchronization mechanism that is independent of hydrodynamic interactions
was recently described in the context of a minimal model swimmer \cite{Friedrich:2012c,Bennett:2013,Polotzek:2013}.
This mechanism crucially relies on the interplay of swimming motion and flagellar beating.

Here, we address the hydrodynamic coupling between the two flagella 
in a model organism for flagellar coordination \cite{Ruffer:1985a,Ruffer:1998a,Polin:2009,Goldstein:2009},
the unicellular green algae \textit{Chlamydomonas}. 
\textit{Chlamydomonas} propels its ellipsoidal 
cell body, which has typical diameter $10\,\micron$,
using a pair of flagella, whose lengths are about $10\,\micron$ \cite{Ruffer:1985a}.
The two flagella beat approximately in a common plane, which is collinear with the long axis of the cell body. 
In that plane, the two beat patterns are nearly mirror-symmetric with respect to this long axis.
The beating of the two flagella of \textit{Chlamydomonas} can synchronize, 
\textit{i.e.} adopt a common beat frequency and a fixed phase relationship 
\cite{Ruffer:1985a,Ruffer:1998a,Polin:2009,Goldstein:2009}.
In-phase synchronization of the two flagella is required for swimming along a straight path \cite{Goldstein:2009}.
The specific mechanism leading to flagellar synchrony is unclear.

Here, we use a combination of realistic hydrodynamic computations and high-speed tracking experiments
to reveal the nature of the hydrodynamic coupling between the two flagella of 
free swimming \textit{Chlamydomonas} cells.
Previous hydrodynamic computations for \textit{Chlamydomonas} 
used either resistive force theory \cite{Tam:2011,Bayly:2011},
which does not account for hydrodynamic interactions between the two flagella,
or computationally intensive finite element methods \cite{Malley:2012}.
We employ an alternative approach and represent the geometry 
of a \textit{Chlamydomonas} cell by spherical shape primitives,
which provides a computationally convenient method that fully accounts for hydrodynamic interactions
between different parts of the cell.
Our theory characterizes flagellar swimming and synchronization by a minimal set of effective degrees of freedom.
The corresponding equation of motion follows naturally from the framework of Lagrangian mechanics,
which was used previously to describe synchronization in a minimal model swimmer \cite{Friedrich:2012c}.
These equations of motion embody the basic assumption that the flagellar beat speeds up
or slows down according to the hydrodynamic friction forces acting on the flagellum.
This assumption is supported by previous experiments, 
which showed that the flagellar beat frequency 
depends on the viscosity of the surrounding fluid
if the concentration of the chemical fuel ATP is sufficiently high \cite{Brokaw:1975}. 
The simple force-velocity relationship for the flagellar beat employed by us
coarse-grains the behavior of thousands of dynein molecular motors that collectively drive the beat.
Similar force-velocity properties have been described for individual molecular motors \cite{Hunt:1994}
and reflect a typical behavior of active force generating systems.

Our theory predicts that any perturbation of synchronized beating results in a significant yawing motion of the cell, 
reminiscent of rocking of the cell body.
This rotational motion imparts different hydrodynamic forces on the two flagella, 
causing one of them to beat faster and the other to slow down.
This interplay between flagellar beating and cell body rocking rapidly restores flagellar synchrony after a perturbation.
Using the framework provided by our theory, we analyze high-speed tracking experiments of swimming cells, 
confirming the proposed two-way coupling between flagellar beating and cell body rocking.

Previous experiments restrained \textit{Chlamydomonas} cells from swimming,
holding their cell body in a micropipette \cite{Ruffer:1998a,Polin:2009,Goldstein:2009}.
Remarkably, flagellar synchronization was observed also for these constrained cells.
This observation seems to argue against a synchronization mechanism that relies on swimming motion.
However, the rate of synchronization observed in these experiments was faster by an order of magnitude
than the rate we predict for synchronization by direct hydrodynamic interactions between the two flagella 
in the absence of any motion.
In contrast, we show that rotational motion with a small amplitude of a few degrees only,
which may result from either a residual rotational compliance of the clamped cell
or an elastic anchorage of the flagellar pair, 
provides a possible mechanism for rapid synchronization,
which is analogous to synchronization by cell body rocking in free-swimming cells.

\section{Results and Discussion}

\subsection{High-precision tracking of confined Chlamydomonas cells}

To study the interplay of flagellar beating and swimming motion, 
we recorded single wild-type \textit{Chlamydomonas reinhardtii} cells swimming in a shallow observation chamber
using high-speed phase-contrast microscopy (1000 fps). 
The chamber heights were only slightly larger than the cell diameter
so that the cells did not roll around their long body axis, 
but only translated and rotated in the focal plane.
This confinement of cell motion to two space dimensions and 
the fact that the approximately planar flagellar beat was parallel to the plane of observation 
greatly facilitated data acquisition and analysis.
From high-speed recordings, we obtained the projected position and orientation of the cell body 
as well as the shape of the two flagella, see figure \ref{fig_dof}A.

In a reference frame of the cell body, each flagellum undergoes periodic shape changes.
To formalize this observation, we defined a flagellar phase variable 
by binning flagellar shapes according to shape similarity, see figure \ref{fig_dof}B.
A time-series of flagellar shapes is represented by a point cloud in an abstract shape space.
This point cloud comprises an effectively one-dimensional shape cycle, 
which reflects the periodicity of the flagellar beat. 
Each shape point can be projected on the centerline of the point cloud.
We define a phase variable $\varphi$ running from $0$ to $2\pi$ that parametrizes this limit cycle 
by requiring that the phase speed $\dot{\varphi}$ is constant for synchronized beating.
Approximately, we determine this parametrization from the condition that the averaged phase speed 
is independent of the location along the limit cycle.
This defines a unique flagellar phase for each tracked flagellar shape.
The width of the point cloud shown in figure \ref{fig_dof}B 
is a measure for the variability of the flagellar beat during subsequent beat cycles. 
We find that the variations of flagellar shapes for the same value of the phase variable
are much smaller than the shape changes during one beat cycle.
For our analysis, we therefore neglect these variations of the flagellar beat.
In this way, we characterize a swimming \textit{Chlamydomonas} cell by $5$ degrees of freedom:
its position $(x,y)$ in the plane, the orientation angle $\alpha$ of its cell body, and the two
flagellar phase variables $\varphi_L$ and $\varphi_R$ for the left and right flagellum, respectively.
Our theoretical description will employ the same $5$ degrees of freedom
and use flagellar shapes tracked from experiment for the hydrodynamic computations.

\subsection{Hydrodynamic forces and interactions}

For a swimming \textit{Chlamydomonas} cell, inertial forces are negligible
(as characterized by a low Reynolds number of $\mathrm{Re}\sim 10^{-3}$ \cite{Malley:2012}),
which implies that the hydrodynamic friction forces exerted by the cell 
depend only on its instantaneous motion \cite{Happel:hydro}.
To conveniently compute hydrodynamic friction forces and hydrodynamic interactions,
we represented the geometry of a \textit{Chlamydomonas} cell by $N=300$ spherical shape primitives, 
see figure~\ref{fig_hydro}A.
The spheres constituting the cell body are treated as a rigid cluster.
For simplicity, we consider free swimming cells and do not include wall effects in our hydrodynamic computations.
Flagellar beating and swimming corresponds to a simultaneous motion of all $300$ spheres of our cell model.
The dependence of the corresponding hydrodynamic friction forces and torques on the velocities of the individual spheres
is characterized by a grand hydrodynamic friction matrix $\G$.
We computed this friction matrix $\G$ using a Cartesian multipole expansion technique \cite{Hinsen:1995}, 
see `Materials and Methods' for details.
Figure~\ref{fig_hydro}C,D shows a sub-matrix that
relates force and velocity components parallel to the long axis of the cell.
The entries of the color matrix depict the force
exerted by any of the flagellar spheres or by the cell body cluster (row index), 
if a single flagellar sphere or the cell body cluster is moved (column index).
The indexing of flagellar spheres is indicated by cartoon drawings of the cell next to the color matrix.
The diagonal entries of this friction matrix are positive and account for the
usual Stokes friction of a single `flagellar sphere' (or of the cell body).
Off-diagonal entries are negative and represent hydrodynamic interactions.
We find considerable hydrodynamic interactions between spheres of the same flagellum,
as well as between each flagellum and the cell body. 
However, interactions between the two flagella are comparably weak.

\subsection{Theoretical description of flagellar beating and swimming}
We present dynamical equations for the minimal set of 5 degrees of freedom shown in figure~\ref{fig_dof}A
to describe flagellar beating, swimming, and later flagellar synchronization in \textit{Chlamydomonas}.
These equations of motion follow naturally from the framework of Lagrangian mechanics of dissipative systems,
which defines generalized forces conjugate to effective degrees of freedom.

Motivated by our experiments, 
we describe the progression through subsequent beat cycles of each of the two flagella 
by respective phase angles $\varphi_L$ and $\varphi_R$, see figure \ref{fig_dof}A.
The angular frequency $\omega_j$ of flagellar beating is given by the time-averaged phase speed 
$\langle\dot{\varphi}_j\rangle$, 
so we can think of the phase speed as the instantaneous beat frequency. 
We are interested in variations of the phase speed that can restore a synchronized state after a perturbation. 
We introduce the key assumption that changes in hydrodynamic friction during the flagellar beat cycle can increase or decrease the phase speed of each flagellum.
Specifically, we assume that for both the left and right flagellum, $j=L,R$, 
the respective flagellar phase speed $\dot{\varphi}_j$ is determined by a balance of 
an active driving force $Q_j$ that coarse-grains the active processes within the flagellum and
a generalized hydrodynamic friction force $P_j$, which depends on $\dot{\varphi}_j$.
Note that in addition to hydrodynamic friction, 
dissipative processes within the flagella may contribute to the friction forces $P_L$ and $P_R$.
We do not consider such internal friction in our description 
as it does not change our results qualitatively.
The hydrodynamic friction forces $P_j$ have to be computed self-consistently for a swimming cell.
We restrict our analysis to planar motion in the `$xy$'-plane and thus consider 
the position $(x,y)$ and the orientation $\alpha$ of the cell body with respect to a fixed laboratory frame,
see figure \ref{fig_dof}A.

Any change of the degrees of freedom $x$, $y$, $\alpha$, $\varphi_L$, $\varphi_R$
results in the dissipation of energy into the fluid at some rate $\mathcal{R}$.
This dissipation rate $\mathcal{R}$ characterizes the mechanical power output of the cell
and plays the role of a Rayleigh dissipation function known in Lagrangian mechanics;
it can be written as
$\mathcal{R}=\dot{x} P_x + \dot{y} P_y + \dot{\alpha} P_\alpha
 + \dot{\varphi}_L P_L+\dot{\varphi}_R P_R$,
which defines the generalized friction forces conjugate to the different degrees of freedom.
The forces $P_L$, $P_R$, $P_\alpha$ are conjugate to an angle and
have physical unit $\mathrm{pN}\,\mu\mathrm{m}$.
We compute the generalized friction forces
using the grand hydrodynamic friction matrix $\G$ introduced above.
In brief, 
the superposition principle of low Reynolds number hydrodynamics relevant for \textit{Chlamydomonas} swimming \cite{Happel:hydro}
implies that the generalized friction forces relate linearly to the generalized velocities,
$P_j = 
\Gamma_{jx}\dot{x}+
\Gamma_{jy}\dot{y}+
\Gamma_{j\alpha}\dot{\alpha}+
\Gamma_{jL}\dot{\varphi}_L+\Gamma_{jR}\dot{\varphi}_R$.
This defines the generalized hydrodynamic friction coefficients $\Gamma_{ji}$,
which are suitable linear combinations of the entries of the grand hydrodynamic friction matrix $\G$, 
see `Materials and Methods'.

The friction force $P_x$ conjugate to the $x$-coordinate of the cell position
represents just the $x$-component of the total force exerted by the cell on the fluid,
an analogous statement applies for $P_y$;
$P_\alpha$ is the total torque associated with rotations around an axis normal to the plane of swimming.
If the swimmer is free from external forces and torques,
we have $P_x=P_y=0$ and $P_\alpha=0$.
Together with the proposed balance of flagellar
friction and driving forces, $P_L=Q_L$ and $P_R=Q_R$,
we have a total of 5 force balance equations, which allow us to solve for
the time derivatives of the 5 degrees of freedom.
We obtain an equation of motion that combines 
swimming and flagellar phase dynamics
\begin{equation}
\label{eq_motion}
(\dot{x},\dot{y},\dot{\alpha},\dot{\varphi}_L,\dot{\varphi}_R)^T =
\Gamma^{-1}(0,0,0,Q_L,Q_R)^T.
\end{equation}
The phase dependence of the active driving forces $Q_j(\varphi_j)$ 
is uniquely specified by the condition 
that the phase speeds should be constant, $\dot{\varphi}_j=\omega_0$,
for synchronized flagellar beating with zero flagellar phase difference $\delta=0$, 
where $\delta=\varphi_L-\varphi_R$.

In essence, this generic description implies that the phase speed of one flagellum 
is determined by hydrodynamic friction forces,
which in turn depend on the swimming motion of the cell.
Since the swimming motion is determined by the beating of \emph{both} flagella,
equation~\ref{eq_motion} effectively defines a feedback loop that couples the two flagellar oscillators.

\subsection{Theory and experiment of Chlamydomonas swimming}

Using the equation of motion (equation \ref{eq_motion}), we can compute the swimming motion of our model cell.
For mirror-symmetric flagellar beating with zero flagellar phase difference $\delta=0$, 
the model cell follows a straight path with an instantaneous velocity that is positive during the effective stroke, 
but becomes negative during a short period of the recovery stroke (figure \ref{fig_swim}A, left panel).
The cell swims \emph{two steps forward, one step back}.
This saltatory motion is also observed experimentally 
by us (figure \ref{fig_swim}A, middle panel)
and others \cite{Racey:1981,Ruffer:1985a,Guasto:2010}.
In our computation, the instantaneous swimming velocity reaches values up to $200\,\mu\mathrm{m}/\mathrm{s}$,
which agrees with experimental measurements for free-swimming cells \cite{Guasto:2010},
but overestimates the observed translational swimming speeds in shallow chambers,
in which wall effects are expected to reduce the speed of translational motion
(compare left and right panels in figure \ref{fig_swim}A).
If the two flagella are beating out of phase, 
the cell will not swim straight anymore, but the cell body yaws, see figure \ref{fig_swim}B.
Cell body yawing is observed experimentally (right panel),
with measured yawing rates that agree well with our computations (left panel). 
The proximity of boundary walls is known
to reduce translational motion,
but to affect rotational motion to a much lesser extent 
for a given distance from the wall \cite{Bayly:2011}.
This is indeed observed in our experiments with cells swimming in shallow chambers:
while the observed translational speed is smaller than predicted (figure \ref{fig_swim}A),
the observed yawing rates are very similar to the predicted ones (figure \ref{fig_swim}B).
The good agreement between theory and experiment for the yawing rate
supports our hydrodynamic computation as well as our description of flagellar beating using a single phase variable.
In the next section, we show that rotational motion is crucial for flagellar synchronization,
while translational motion is less important.

\subsection{Theory of flagellar synchronization by cell body yawing}

We now demonstrate how yawing of the cell body leads to flagellar synchronization.
We first examine the flagellar phase dynamics after a perturbation of in-phase flagellar synchrony.
Figure \ref{fig_sync}A shows numerical results for a free swimming cell
obtained from solving the equation of motion (equation \ref{eq_motion}).
The initial flagellar asynchrony causes a yawing motion of the model cell,
which is characterized by periodic changes of the cell's orientation angle $\alpha(t)$.
The phase difference $\delta$ between the left and right flagellum decays approximately exponentially as
$\delta(t)\sim\exp(-\lambda t/T)$
with a rate constant $\lambda$ [measured in beat periods $T=2\pi/\omega_0$]
that will serve as a measure of the strength of synchronization.

To mimic experiments where external forces constrain cell motion,
we now consider the idealized case of a cell that cannot translate,
while cell body yawing is constrained by an elastic restoring force $Q_\alpha=-\kyaw\alpha$.
Again, the two flagella synchronize in-phase, provided some residual cell body yawing is allowed, see figure \ref{fig_sync}B.
In the absence of an elastic restoring force ($\kyaw=0$),
when the model cell cannot translate, but can still freely rotate,
its yawing motion and synchronization behavior is very similar to the case of a free swimming cell
that can rotate \textit{and} translate.
For a fully clamped cell body, however,
the synchronization strength is strongly attenuated,
and is solely due to the direct hydrodynamic interactions between the two flagella.
In this case of synchronization by hydrodynamic interactions, 
the time-constant for synchronization is decreased approximately 20-fold compared to the case of free swimming.
These numerical observations point at a crucial role of cell body yawing for flagellar synchronization.
The underlying mechanism of synchronization can be explained as follows.
For in-phase synchronization, the flagellar beat is mirror-symmetric and the cell swims along a straight path.
If, however, the left flagellum has a small head-start during the effective stroke,
this causes a counter-clockwise rotation of the cell, see figure~\ref{fig_swim}B.
This cell body yawing increases (decreases) the hydrodynamic friction encountered by the left (right) flagellum, 
causing the left flagellum to beat slower and the right one to beat faster. 
As a result, flagellar synchrony is restored.

\begin{figure}[hp]
\includegraphics[width=8.5cm]{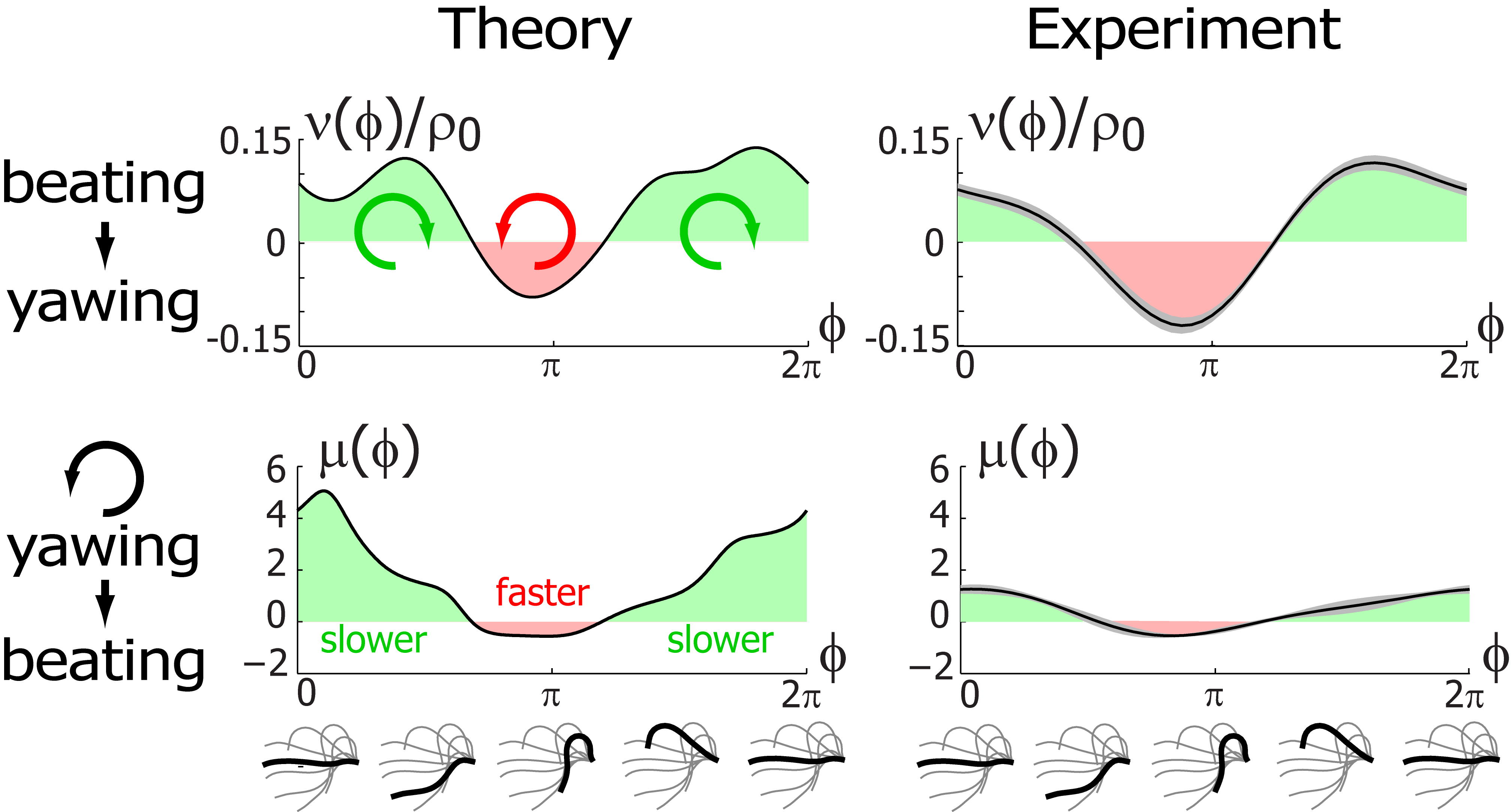}
\caption{
\label{fig_yaw}
Flagellar beating and cell body yawing are coupled in a bidirectional way.
\textbf{Theory (left):} 
In our theory, the beat of the left flagellum generates a torque, 
which, in the absence of the right flagellum, has to be counterbalanced by a yawing motion of cell body, see eqn.~\ref{eq_yaw3}.
This effect is quantified by the coupling function $\nu(\varphi)$
shown in the upper panel, normalized here by $\rho_0=\langle\rho\rangle$:
the effective stroke ($\varphi_L\sim 0^\circ$) of the left flagellum causes the cell to yaw clockwise.
Conversely, 
yawing of the cell
changes the hydrodynamic friction force that opposes the flagellar beat,
which, in our theory, speeds up or slows down the beat, see eqn.~\ref{eq_yaw1}.
This effect is quantified by the coupling function $\mu(\varphi)$
shown in the lower panel:
a counter-clockwise yawing during the effective stroke of the left flagellum slows down its beat.
The coupling of beating and yawing allows for flagellar synchronization in a free swimming cell.
\textbf{Experiment (right):}
By fitting eqns.~\ref{eq_yaw1} and \ref{eq_yaw3} to experimental time series data, 
we can recover the coupling functions $\mu(\varphi)$ and $\nu(\varphi)/\rho_0$ 
(1 cell, $ n=5$ time series of $0.5\,\mathrm{s}$ duration, gray regions denote mean$\pm$s.e.).
}
\end{figure}

Next, we present a formalized version of this argument using a reduced equation of motion.
We thus arrive at a simple theory for biflagellar synchronization, 
which will later allow for quantitative comparison with experiments.
As in figure \ref{fig_sync}B,
we assume that the cell is constrained such that it cannot translate ($\dot{x}=\dot{y}=0$).
The cell can still yaw, possibly being subject to an elastic restoring force $Q_\alpha=-\kyaw\alpha$.
This leaves only three degrees of freedom: $\varphi_L$, $\varphi_R$, and $\alpha$.
Neglecting direct hydrodynamic interactions between the flagella, 
we can reduce the equations of motion for a clamped cell 
(equation \ref{eq_motion} with constraint $\dot{x}=\dot{y}=0$)
to a set of three coupled equations for the three remaining degrees of freedom
\begin{align}
\label{eq_yaw1}
\dot{\varphi}_L &= \omega_L - \mu(\varphi_L)\dot{\alpha} , \\
\label{eq_yaw2}
\dot{\varphi}_R &= \omega_R + \mu(\varphi_R)\dot{\alpha} , \\
\label{eq_yaw3}
\kyaw\,\alpha+\rho(\varphi_L,\varphi_R)\,\dot{\alpha}
&= -\nu(\varphi_L)\dot{\varphi}_L+\nu(\varphi_R)\dot{\varphi}_R.
\end{align}
The coupling function $\mu$ in equation \ref{eq_yaw1}
characterizes the effect of cell body yawing on the flagellar beat as detailed below,
while $\nu$ describes how asynchronous flagellar beating results in yawing;
$\rho$ is the hydrodynamic friction coefficient for yawing of the whole cell.
The coupling functions $\mu$, $\nu$, and $\rho$ can be computed using our hydrodynamic model%
\footnote{Specifically, 
$\mu(\varphi) = \Gamma_{L\alpha}(\varphi,\varphi) / \Gamma_{LL}(\varphi,\varphi)$,
$\nu(\varphi) = \Gamma_{\alpha L}(\varphi,\varphi)$,
$\rho(\varphi_L,\varphi_R)= \Gamma_{\alpha\alpha}(\varphi_L,\varphi_R)$.
For simplicity, the active flagellar driving forces 
were approximated as 
$Q_L=\omega_L\Gamma_{LL}$ and $Q_R=\omega_R\Gamma_{RR}$
for constrained translational motion.}.
Their dependence on the flagellar phase is shown in figure \ref{fig_yaw} (left panels).
The physical significance of equations \ref{eq_yaw1}-\ref{eq_yaw3} can be explained as follows:
Equation \ref{eq_yaw1} implies that 
during the effective stroke of the left flagellum ($\varphi\sim 0^\circ$),
a counter-clockwise rotation of the whole cell 
slows down the flagellar beat, while a clockwise rotation speeds it up (figure \ref{fig_yaw}B, $\mu>0$).
Equation \ref{eq_yaw2} implies the converse for the right flagellum.
During the recovery stroke ($\varphi\sim 180^\circ$), 
the effect is opposite and a counter-clockwise rotation of the cell
would speed up the beat of the left flagellum ($\mu<0$).
Equation \ref{eq_yaw3} states that flagellar beating causes the cell body to yaw:
if the right flagellum were absent, the model cell rotates clockwise ($\dot{\alpha}<0$)
during the effective stroke of the left flagellum (figure \ref{fig_yaw}A, $\nu>0$), 
and counter-clockwise during its recovery stroke ($\nu<0$).
This swimming behavior is observed for uniflagellar mutants \cite{Bayly:2011}.
For synchronized beating of the two flagella, the right-hand side of equation \ref{eq_yaw3}
cancels to zero and the model cell swims straight.
For asynchronous flagellar beating with a finite phase-difference 
$\delta=\varphi_L-\varphi_R$,
the phase-dependence of the coupling function $\nu(\varphi)$ results
in an imbalance of the torques generated by the left and right flagellum, 
respectively, which is balanced by a rotation of the whole cell.
A similar situation arises for an elastically anchored flagellar pair
as detailed in the Supporting Information (SI).
In that scenario, the flagellar pair pivots as a whole when the two flagella beat asynchronously,
which causes rapid synchronization analogous to synchronization by cell body yawing considered here.

We study the dynamical system given by equations \ref{eq_yaw1}-\ref{eq_yaw3}
after a small perturbation of the synchronized state at $t=0$
with initial flagellar phase difference $0<\delta(0)\ll 1$.
For simplicity, we assume intrinsic beat frequencies, $\omega_L=\omega_R=\omega_0$.
For small initial perturbations, the flagellar phase difference
will either decay or grow exponentially, if we average its dynamics over a beat cycle,
$\delta(t)\approx\delta(0)\exp(-\lambda t/T)$.
A positive value of the synchronization strength $\lambda$ implies that the perturbation decays,
and that the in-phase synchronized state with $\delta=0$ is stable.
The synchronization strength is given by $\lambda=-\int_0^T dt\,\dot{\delta}/\delta$.
In the limit of a small elastic constraint, we find (see SI for details)
\begin{equation}
\label{eq_lyaw}
\lambda 
= - \oint_0^{2\pi}\!\! d\varphi\,\, \frac{2\mu(\varphi)\nu'(\varphi)}{\rho(\varphi,\varphi)-2\mu(\varphi)\nu(\varphi)}
\text{ for }\kyaw\ll\rho\omega_0,
\end{equation}
where a prime denotes differentiation with respect to $\varphi$.
Using the coupling functions $\mu$, $\nu$, $\rho$ computed above,
we obtain $\lambda>0$, which implies stable in-phase synchronization, see figure~\ref{fig_sync}.

In the case of a stiff elastic constraint, 
we obtain a different result for the synchronization strength $\lambda$
\begin{equation}
\label{eq_lelastic}
\lambda = - \oint_0^{2\pi}\!\! d\varphi\,\, \frac{\mu(\varphi)\nu''(\varphi)}{\kyaw/\omega_0}
\text{ for }\kyaw\gg\rho\omega_0.
\end{equation}

Synchronization in the absence of an elastic restoring force as characterized by equation \ref{eq_lyaw},
and synchronization involving a strong elastic coupling as characterized by equation \ref{eq_lelastic}
show interesting differences, which relate to the fact that in the first case  
the flagellar phase dynamics depends only on the yawing rate $\dot{\alpha}$, but not on $\alpha$ itself.
The difference between these two synchronization mechanisms is best illustrated 
in a special case, in which both the ratio $\sigma=\mu/\nu$ and $\rho$ are constant.
A constant $\sigma$ correspond to an active flagellar driving force that does not depend on the flagellar phase,
whereas for constant $\rho$, the angular friction for yawing would not depend on the flagellar configuration.
In the limit of a stiff elastic constraint, $\kyaw\gg\rho\omega_0$, 
we readily find $\lambda=-\sigma\omega_0\oint\! \nu\nu''/\kyaw=\sigma\omega_0\oint\! (\nu')^2/\kyaw>0$,
which indicates stable in-phase synchronization.
In the limit of a weak elastic constraint, $\kyaw\ll\rho\omega_0$, however,
the integral on the right-hand side of equation~\ref{eq_lyaw} evaluates to zero,
which implies that synchronization does not occur.
Hence, synchronization in the absence of an elastic restoring force requires 
that either $\mu/\nu$ or $\rho$ depend on the flagellar phase.

For our realistic \textit{Chlamydomonas} model,
$\mu$ and $\nu$ differ (figure \ref{fig_yaw}A),
and also $\rho$ is not constant (figure \ref{fig_gh} in the SI).
This allows for rapid synchronization also in the absence of elastic forces.
Previous work on synchronization in minimal systems showed that elastic restoring forces can facilitate synchronization
\cite{Reichert:2005,Niedermayer:2008}.
Here, we have shown that elastic forces can increase the synchronization strength (figure \ref{fig_sync}),
but they are not required for flagellar synchronization in swimming \textit{Chlamydomonas} cells,
even if hydrodynamic interactions are neglected.

Our discussion of flagellar synchronization can be extended to the case, 
where the intrinsic beat frequencies of the two flagella do not match.
If the frequency mismatch 
$|\omega_L-\omega_R|$
is small compared to the inverse time-scale of synchronization $\lambda/T$,
a general result implies that the two flagellar oscillators will still synchronize \cite{Pikovsky:synchronization}.
For a frequency mismatch that is too large, however, 
the two flagella will display phase drift with monotonously increasing phase difference.

\subsection{Experiments show coupling of beating and yawing}
We reconstructed the coupling functions $\mu(\varphi)$ and $\nu(\varphi)$ 
between beating and yawing from experimental data
using the theoretical framework developed in the previous section.
In brief, 
(1) we extracted the instantaneous yawing rate $\dot{\alpha}$ 
and flagellar phase speeds $\dot{\varphi}_L$ and $\dot{\varphi}_R$ from high-speed videos of swimming \textit{Chlamydomonas} cells,
(2) we represented the coupling functions by a truncated Fourier series, and 
(3) we obtained the unknown Fourier coefficients by linear regression using equations \ref{eq_yaw1}-\ref{eq_yaw3}.
The high temporal resolution of our imaging enabled us 
to accurately determine phase speeds as time-derivatives of flagellar phase angle data.
Figure \ref{fig_yaw}B displays averaged coupling functions obtained by fitting for a typical \textit{Chlamydomonas} cell.
We find a significant coupling between flagellar phase speeds and yawing rates,
which are in good qualitative agreement with the theoretical predictions.
Figure \ref{fig_fits} in the SI shows fits for 5 more cells.

For the experimental conditions used,
we commonly observed cells that displayed a large frequency mismatch between the two flagella.
In the cells selected for analysis, this frequency mismatch exceeded $30\%$.
This large frequency mismatch caused flagellar phase drift, 
which resulted in pronounced cell body yawing 
and enabled us to accurately measure the coupling of yawing and flagellar beating.
Experiments were done using either white light illumination, which gave maximal image quality,
or red light illumination, which reduces a possible phototactic stimulation of the cells.

The observed modulation of flagellar phase speed according to the rate of yawing
is consistent with a force-velocity dependence of flagellar beating,
for which the speed of the beat decreases, if the hydrodynamic load increases.
We propose that a similar load characteristic of the flagellar beat 
holds also in cases of small frequency mismatch,
where it allows for flagellar synchronization.

\section{Conclusion and Outlook}

We have presented a theory on the hydrodynamic coupling
underlying flagellar synchronization in swimming \textit{Chlamydomonas} cells.
We have shown that direct hydrodynamic interactions between the two flagella as
considered in refs.\ \cite{Gueron:1999,Vilfan:2006,Niedermayer:2008} give only
a minor contribution to the computed synchronization strength and are unlikely
to account for the rapid synchronization observed in experiments 
\cite{Ruffer:1985a,Ruffer:1998a,Polin:2009,Goldstein:2009}.
In contrast, 
rotational motion of the swimmer caused by asynchronous beating
imparts different hydrodynamic friction forces on the two flagella, 
which rapidly brings them back in tune:
\textit{Chlamydomonas} \emph{rocks to get into synchrony}.

Using high-speed tracking experiments,
we could confirm the two-way coupling between flagellar beating and cell body yawing predicted by our theory.
The striking reproducibility of our fits for the corresponding coupling functions
and their favorable comparison to our theory 
is highly suggestive for a regulation of flagellar phase speed by hydrodynamic friction forces that depends on rotational motion.
Thus, coupling of flagellar beating and cell body yawing provides a strong candidate
for the mechanism that underlies flagellar synchronization of swimming \textit{Chlamydomonas} cells.
A similar mechanism may account for synchronization in isolated flagellar pairs \cite{Hyams:1975}, 
see figure \ref{fig_sync_isolated_pair}.

To explain a previously observed synchronization for cells held in a micropipette 
\cite{Ruffer:1998a,Polin:2009,Goldstein:2009},
we propose a finite clamping compliance that still allows for residual cell body yawing
with an amplitude of a few degrees, which is sufficient for rapid synchronization.
Alternatively, a compliant basal anchorage of the flagellar pair or bending deformations of the elastic cell body
would allow for flagellar synchronization by a completely analogous mechanism.
The simple theory for biflagellar synchronization by rotational motion presented in this manuscript
(equations~\ref{eq_yaw1}-\ref{eq_yaw3}) 
is generic and applies also to the pivoting motion of an elastically anchored flagellar pair 
as shown in the Supporting Information.
From the observed value $\lambda=0.3$ for the synchronization strength in clamped cells \cite{Goldstein:2009},
we estimate a rotational stiffness of $k\sim 10^4\,\mathrm{pN}\,\micron$ for either of these two cases.

Finally, the coupling of two phase oscillators by a third degree of freedom,
in this case rotational motion, could allow for synchronization also in other contexts.
For example, one may consider that synchronization in ciliar arrays \cite{Sanderson:1981}
is mediated by an elastic coupling through the matrix
with elastic deformations playing the role of the third degree of freedom.

\section{Hydrodynamic computation of swimming \textit{Chlamydomonas}}
We represent a Chlamydomonas cell by an ensemble of $ 300$ spheres of radius 
$ a=0.25\,\micron$, see Figure \ref{fig_hydro}A, and use a freely available hydrodynamic library 
based on a Cartesian multipole expansion technique \cite{Hinsen:1995}
to compute the grand hydrodynamic friction matrix $\mathrm{G}$ \cite{Happel:hydro} for this ensemble of spheres. 
We assume a rigid cell body, 
and hence that the spheres constituting the cell body move as a rigid unit,
which results in $ n=2\cdot 14+1$ independently moving objects.
The matrix $\mathrm{G}$ has dimensions $ 6n\times 6n$ and 
relates the components of the translational and rotational velocities, 
$\v_i$ and $\Omega_i$, of each of the $ n$ objects 
to the hydrodynamic friction forces and torques, $\mathrm{F}_j$ and $\mathrm{T}'_j$, exerted by the $ j$-th object on the fluid,
$(F_{1x},F_{1y},F_{1z},T'_{1x},T'_{1y},T'_{1z},F_{2x},F_{2y},\ldots,T'_{ny},T'_{nz}) = 
\mathrm{G} \dot{\q}_0$
with
$\dot{\q}_0=(v_{1x},v_{1y},v_{1z},\Omega_{1x},\Omega_{1y},\Omega_{1z},v_{2x},v_{2y},\ldots,\Omega_{ny},\Omega_{nz})$.
Figure~\ref{fig_hydro}C,D shows the sub-matrix 
$ G_{ij,yy}=G_{6i-4,6j-4}$,
which relates $ y$-components of velocities and $ y$-components of hydrodynamic forces.
The reduced friction matrix $\Gamma$ for a set of $ m$ effective degrees of freedom $ q$ 
is computed from $\mathrm{G}$ as
$ \Gamma=\mathrm{L}^T \mathrm{G} \mathrm{L}$ with $ 6n\times m$ transformation matrix 
$ L_{ij}=\partial{\dot{q}_{0,i}}/\partial{\dot{q}_j}$,
where $ \q=(x,y,\alpha,\varphi_L,\varphi_R)$ \cite{Friedrich:2012c}.
Initial tests confirmed that the friction matrix of only the cell body 
gave practically the same result as the analytic solution for the enveloping spheroid; 
similarly the computed friction matrix of only a single flagellum matched the prediction of resistive force
theory \cite{Happel:hydro}. 

\section{Imaging \textit{Chlamydomonas} swimming in a shallow observation chamber} %
Cell Culture: Chlamydomonas reinhardtii cells (CC-125 wild type mt+ 137c, R.P. Levine via N.W. Gillham, 1968) were grown in 300 ml TAP+P buffer \cite{Gorman:1965} (with 4 x phosphate) at $ 24^\circ$ C for 2 days under conditions of constant illumination (2x75 W, fluorescent bulb) and constant air bubbling to a final density of $ 10^6$ cells/ml. 

High Speed Video Microscopy: 
An assay chamber was made of pre-cleaned glass and sealed using Valap, a 1:1:1 mixture of lanolin, paraffin, and petroleum jelly, heated to $ 70^\circ\,\mathrm{C}$. 
The surface of that chamber was blocked using casein solution (solution of casein from bovine milk 2 mg/ml, for 10 min) prior to the experiment. 
Single, non interacting cells were visualized using phase contrast microscopy set up on a Zeiss Axiovert 100 TV Microscope using a 63x Plan-Apochromat NA1.4 PH3 oil lens in combination with an 1.6x tube lens and an oil phase-contrast condenser NA 1.4. 
The sample was illuminated using a 100 W Tungsten lamp. 
For red-light imaging, an e-beam driven luminescent light pipe (Lumencor, Beaverton, USA) with spectral range of 640-657 nm 
and power of 75 mW was used. 
The sample temperature was kept constant at $ 24^\circ\,\mathrm{C}$ using an objective heater (Chromaphor, Oberhausen, Germany). 
For image acquisition, an EoSens Cmos high-speed camera was used. 
Videos were acquired at a rate of 1000 fps with exposure times of 1 ms (white light) and 0.6 ms (red light).  
Finally, cell positions and flagellar shapes were tracked using custom-build Matlab software, see SI for details.





\forcenewpage

\section{Figures}

\begin{figure}[hp]
\includegraphics[width=8.5cm]{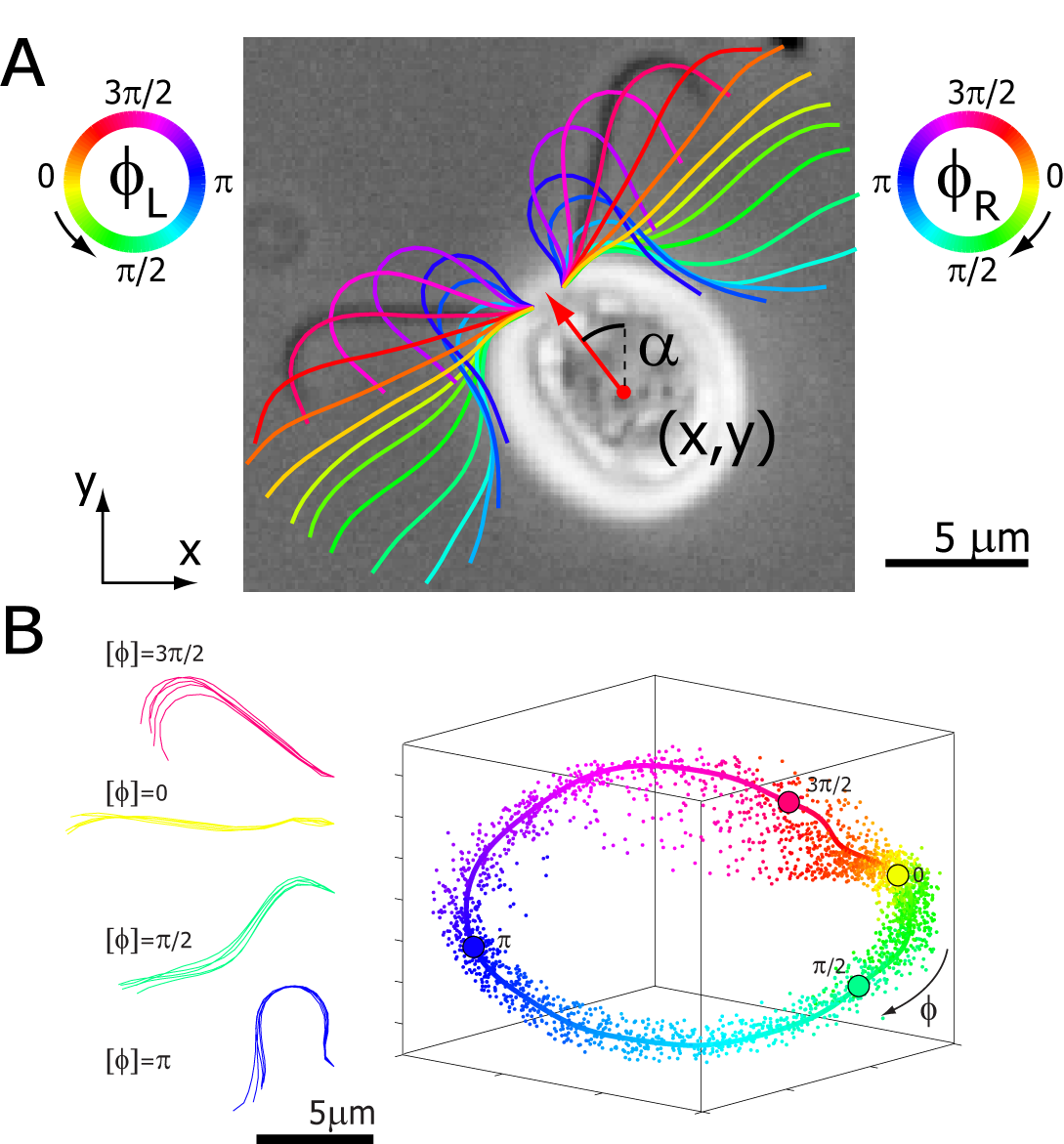}
\caption{
\label{fig_dof}
Five degrees of freedom for \textit{Chlamydomonas}.
\textbf{A.}
In our experiments, conducted in shallow observation chambers,
\textit{Chlamydomonas} cells swim in a plane.
At each time, the position and orientation of the cell body is characterized
by its center position $ (x,y)$ and the angle $\alpha$ of its long axis with respect to 
the laboratory frame.
The beating of each flagellum is characterized by a single periodic phase variable,
$\varphi_L$ and $\varphi_R$ for the left and right flagellum, respectively.
The flagellar shapes shown in different colors were 
tracked from high-speed recordings and correspond to a time-difference of $ 2\,\mathrm{ms}$.
This beat pattern was used for all computations.
\textbf{B.}
Binning of tracked flagellar shapes according to shape similarity
defines a flagellar phase angle as shown on the left.
More precisely,  
we employed a nonlinear dimensionality reduction technique as 
specified in the Supporting Information to represent
each tracked planar flagellar shape as a point in an abstract shape space.
This representation reveals the periodicity of the flagellar beat
and supports our description of the flagellar beat as a fixed sequence
of flagellar shapes parameterized by a single phase variable $\varphi$.
}
\end{figure}

\forcenewpage

\begin{figure}[hp]
\includegraphics[width=8.5cm]{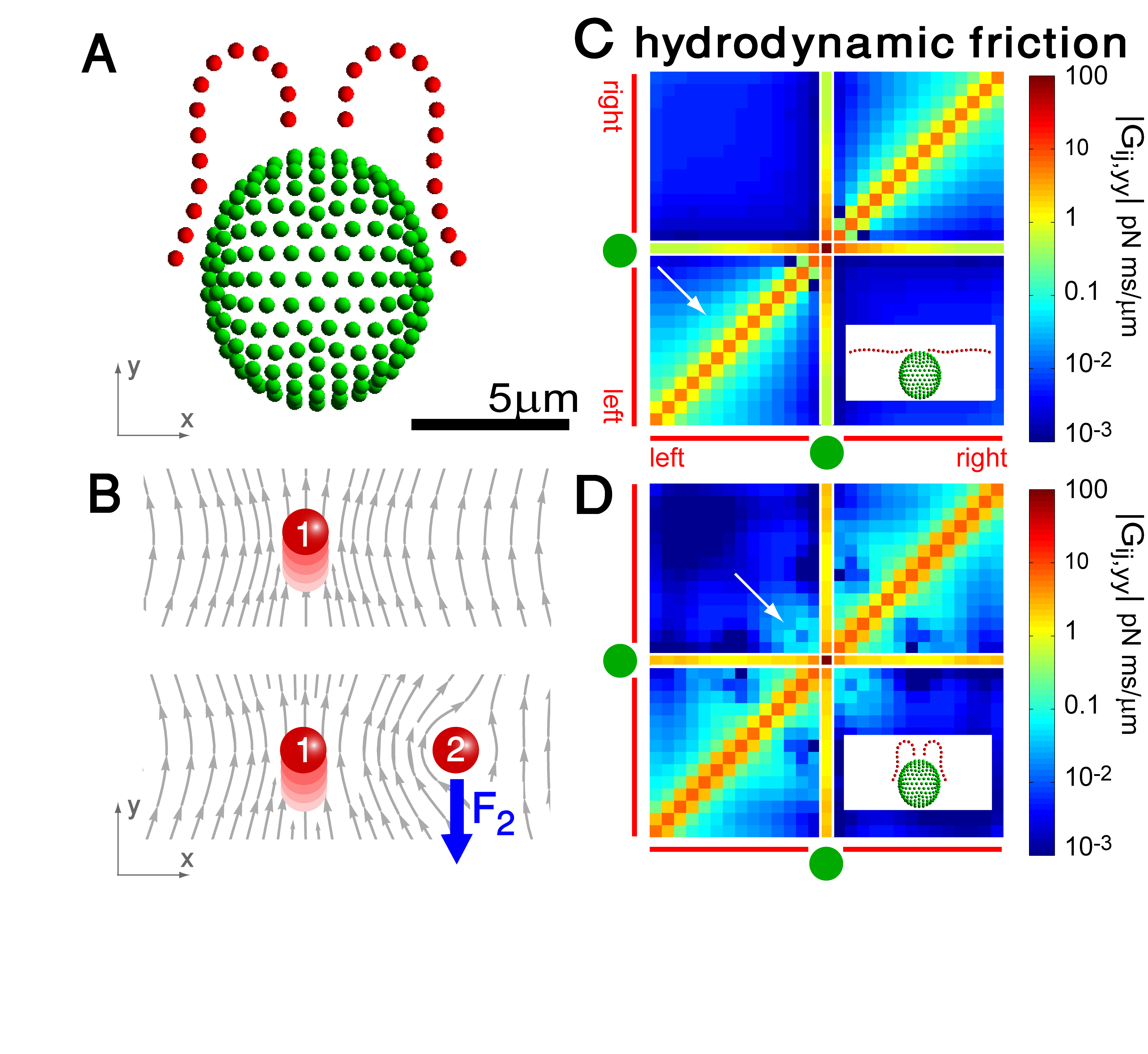}
\caption{
\label{fig_hydro}
Hydrodynamic interactions between the two flagella are weak.
\textbf{A.}
Model \textit{Chlamydomonas} cell represented by an ensemble of $ 300$ spheres
used to compute hydrodynamic friction forces at low Reynolds numbers.
In our calculations, the model cell was assumed to be far from any surfaces.
\textbf{B.} 
Illustration of hydrodynamic interactions between spheres.
A single sphere (labeled $ 1$) moving with velocity $ v_{1y}>0$
along the $ y$-axis will drag fluid alongside and thus 
exert a total hydrodynamic friction force $ F_{1y}=G_{11,yy}v_{1y}>0$ on the fluid.
If a second sphere (labeled $ 2$) is held fixed close to the first one, 
it will locally slow down this fluid flow.
The force $ \mathrm{F}_2$ required to hold the second sphere equals the force exerted by this sphere on the fluid;
its $ y$-component $ F_{2y}=G_{21,yy}v_{1y}<0$ 
defines a friction coefficient $ G_{21,yy}$ that characterizes hydrodynamic interactions between the two spheres.
\textbf{C.}
Hydrodynamic interactions between different parts of the model cell.
Analogous to panel B, one defines a matrix $ G_{ij,yy}$
of hydrodynamic friction coefficients 
for the ensemble of $ 2\cdot 14$ flagellar spheres 
and the rigid sphere cluster constituting the cell body that together represent a \textit{Chlamydomonas} cell (inset).
Each column of the color coded matrix shows the magnitude of hydrodynamic friction
exerted by a flagellar sphere (or the cell body),
if a single sphere or the cell body is moved 
parallel to the long cell body axis.
Off-diagonal entries characterize hydrodynamic interactions,
which are particularly pronounced along a single flagellum (white arrow),
or between one flagellum and the cell body (central column).
Hydrodynamic interactions between the two flagella are very weak
and partly screened by the cell body.
\textbf{D.}
Same as in panel C, but for a recovery stroke configuration.
There are weak hydrodynamic interactions between the proximal segments
of the two flagella (white arrow).
All friction coefficients shown scale with the viscosity of the fluid,
which was taken as the viscosity of water at $ 20^\circ$ Celsius,
$\eta=1\,\mathrm{pN\,ms}/\micron^2$.
}
\end{figure}

\forcenewpage

\begin{figure}[hp]
\includegraphics[width=8.5cm]{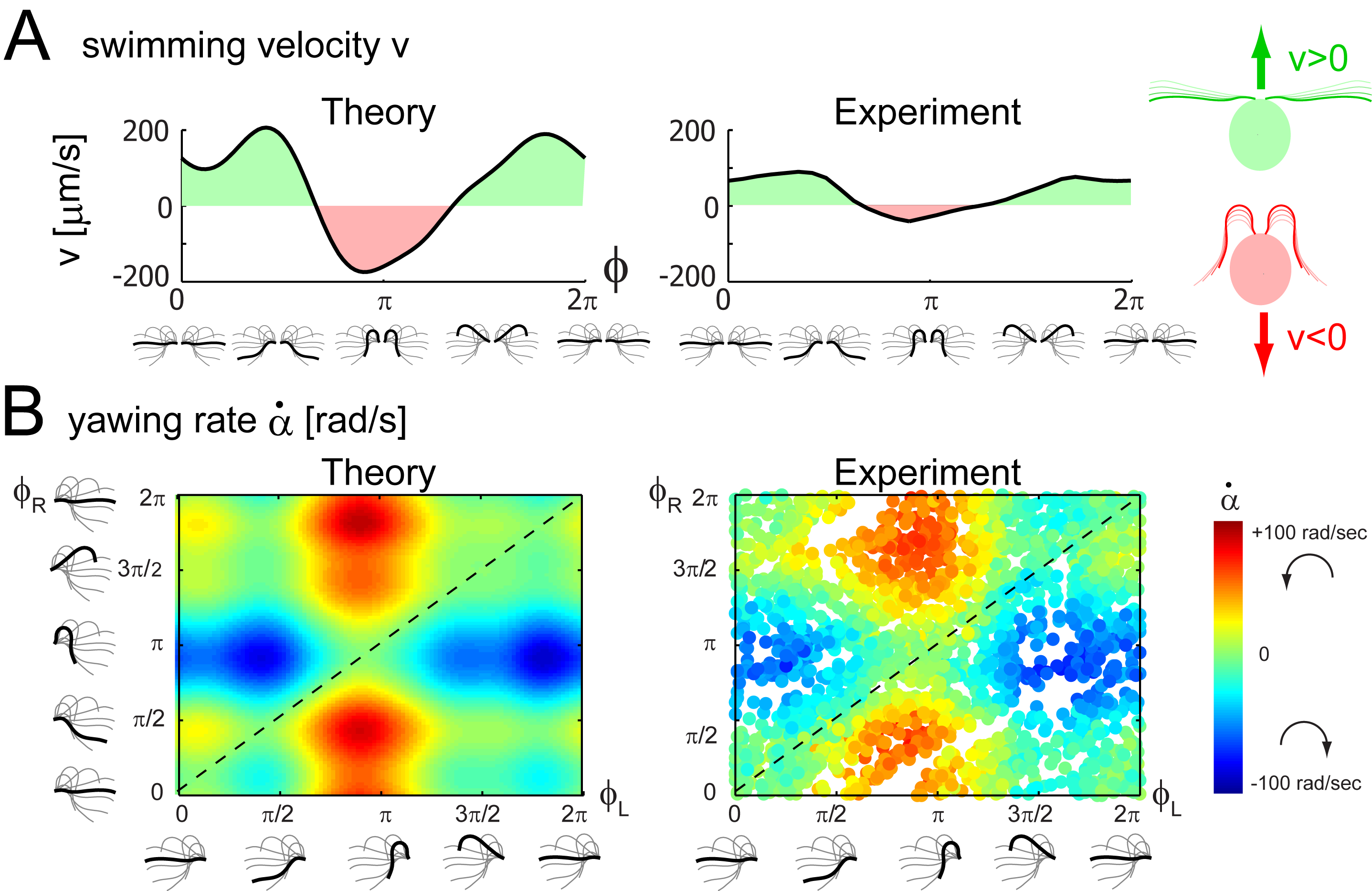}
\caption{
\label{fig_swim}
\textbf{A.}
For synchronized flagellar beating, we compute
saltatory forward swimming with positive instantaneous velocity during effective stroke beating,
and a backward motion during the recovery stroke (left panel);
this behavior is summarized by cartoon drawings to the very right.
A typical experimental velocity profile 
of a \textit{Chlamydomonas} cell in a shallow observation chamber 
measured during a cycle of synchronized beating is shown for comparison in the middle panel.
\textbf{B.}
Flagellar asynchrony causes cell body yawing, both in theory and experiment.
Shown is the instantaneous rotation rate $\dot{\alpha}$ of the cell body in color code 
as a function of the respective phase of the two flagella.
For in-phase synchronized flagellar beating (dashed line), 
the cell body does not rotate (green).
For out-of-phase flagellar beating, however, 
we find significant cell body rocking (blue: clockwise, red: counter-clockwise).
}
\end{figure}

\forcenewpage

\begin{figure}[hp]
\includegraphics[width=8.5cm]{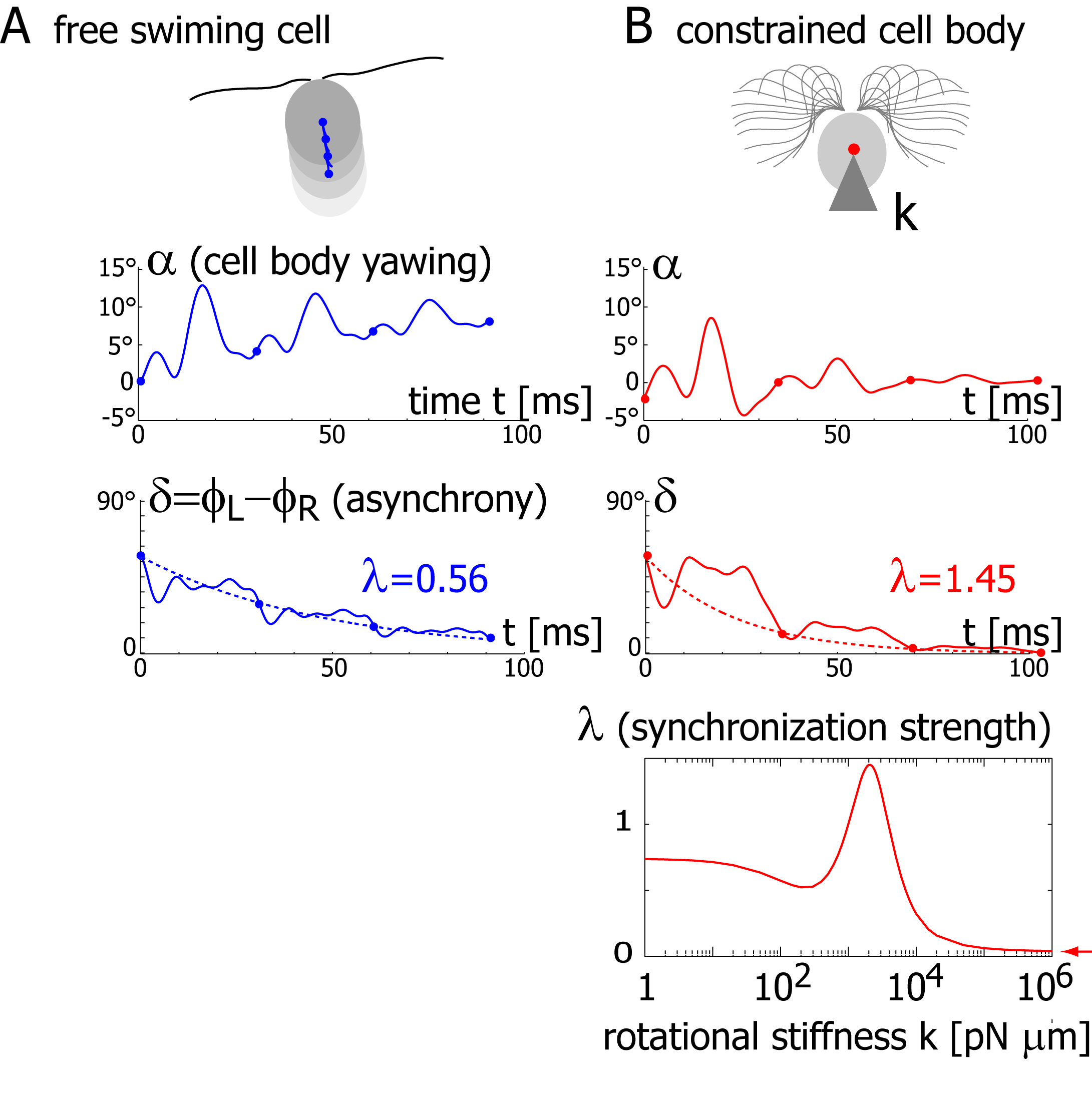}
\caption{
\label{fig_sync}
Flagellar synchronization by cell body yawing.
\textbf{A.}
For a free swimming cell (top panel),
the equation of motion \ref{eq_motion} predicts a yawing motion of the cell body characterized by $\alpha(t)$
if the two flagella are initially out of synchrony (middle panel).
The flagellar phase difference $\delta$ is found to decrease with time (lower panel, solid line),
approximately following an exponential decay $\sim\exp(-\lambda t/T)$ (dotted line),
where $ T$ is the period of the flagellar beat and $\lambda$ defines
a dimensionless synchronization strength.
Thus, in-phase synchronized beating is stable with respect to perturbations.
Dots mark the completion of a full beat cycle of the left flagellum.
\textbf{B.}
To mimic experiments where external forces constrain cell motion,
we simulated the idealized case of a cell that cannot translate,
while cell body yawing is constricted by an elastic restoring torque $ Q_\alpha=-\kyaw\alpha$
that acts at the cell body center (upper panel).
Again, the two flagella synchronize ($ 3^\mathrm{rd}$ panel) 
with a synchronization strength $\lambda$
that can become even larger than in the case of a free swimming
as shown here for $\kyaw=2\cdot 10^3\,\mathrm{pN}\,\mu\mathrm{m}$,
which is close to the rotational stiffness
for which the synchronization strength $\lambda$ is maximal (lowest panel).
For very large clamping stiffness $\kyaw$, 
the cell body cannot move and the synchronization strength $\lambda$
attenuates to a basal value $\lambda\approx 0.03$, which arises solely from
direct hydrodynamic interactions between the two flagella (arrow).
Parameters: $2\pi/\omega_0=30\,\mathrm{ms}$.
}
\end{figure}



\forcenewpage

\forcenewpage

\section{Supporting Information}

\renewcommand{\thefigure}{S\arabic{figure}}
\setcounter{figure}{0}
\renewcommand{\theequation}{S\arabic{equation}}
\setcounter{equation}{0}

\subsection{Image analysis} %
\begin{figure}
\includegraphics[width=8.5cm]{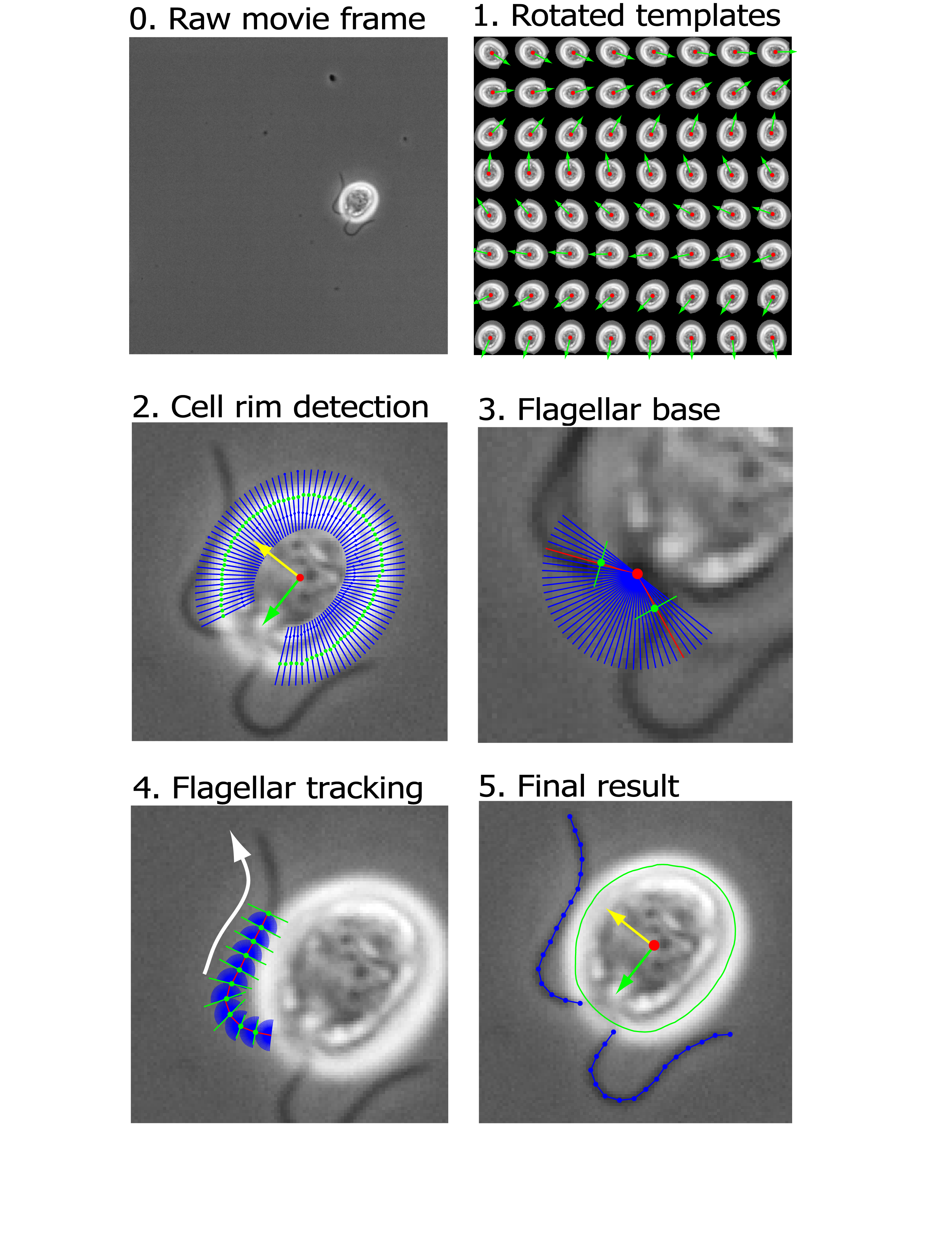}
\caption{
\label{fig_tracking}
Image analysis pipeline used to automatically track 
planar cell position and orientation as well as flagellar shapes 
in high-speed movies of swimming \textit{Chlamydomonas} cells.
\textbf{0.} A typical movie frame 
\textbf{1.} Rotated template images used for a cross-correlation analysis
to estimate cell position and orientation in a movie frame.
\textbf{2.} 
The cell body outline was tracked by detecting intensity maxima (green) 
of line scans along rays (shown in blue), which emanate from the putative cell body center. 
From the cell body outline, we obtain refined estimates for cell position and orientation.
\textbf{3.} The position of the flagellar base was then determined using a fan of line-scans (along the blue lines),
followed by a line-scan (green) in a direction perpendicular to the maximal intensity direction (red).
\textbf{4.} Finally, flagellar shapes were tracked in a successive manner using 
similar combinations of line-scans as in step 3.
}
\end{figure}

High-speed movies were analyzed using custom-made Matlab software (The MathWorks Inc., Natnick, MA, USA);
our image analysis pipeline is illustrated in figure~\ref{fig_tracking}:
In a first step, estimates for position and orientation of the cell body in a movie frame
were obtained by a cross-correlation analysis using rotated template images.
In a second step, these position and orientation estimates were 
refined by tracking the bright phase halo surrounding the cell.
The first and second area moments of the cell rim 
provide accurate estimates for the center of the cell body and its long orientation axis:
While the tracking precision of the first step amounts to $<500\mathrm{nm}$ 
for the position and a few degrees for the orientation, 
these values are reduced to $<50\mathrm{nm}$ and $<0.5^\circ$ after the second step, respectively.
Special care was taken to reduce any potential bias of the flagellar phase on the cell body tracking;
for example, the cell rim close to the flagellar bases was obtained by interpolation instead of direct tracking.
The flagellar base is visible as a continuous, parabola-shaped curve that connects the proximal ends of the two flagella;
tracking of this flagellar base was done by a combination of line-scans and local fitting of a Gaussian line model (step 3).
Flagella were tracked by advancing along their length 
using exploratory line-scans in a successive manner (step 4).
Flagellar tracking can be refined by local fitting of a Gaussian line model.
A movie consisting of thousand frames can be analyzed in an automated manner within 10 hours on a standard PC.
Movies from red light illumination conditions were of lower quality and required manual correction
of the automated tracking results for each frame.

\subsection{Flagellar shape analysis}

\begin{figure}
\includegraphics[width=8.5cm]{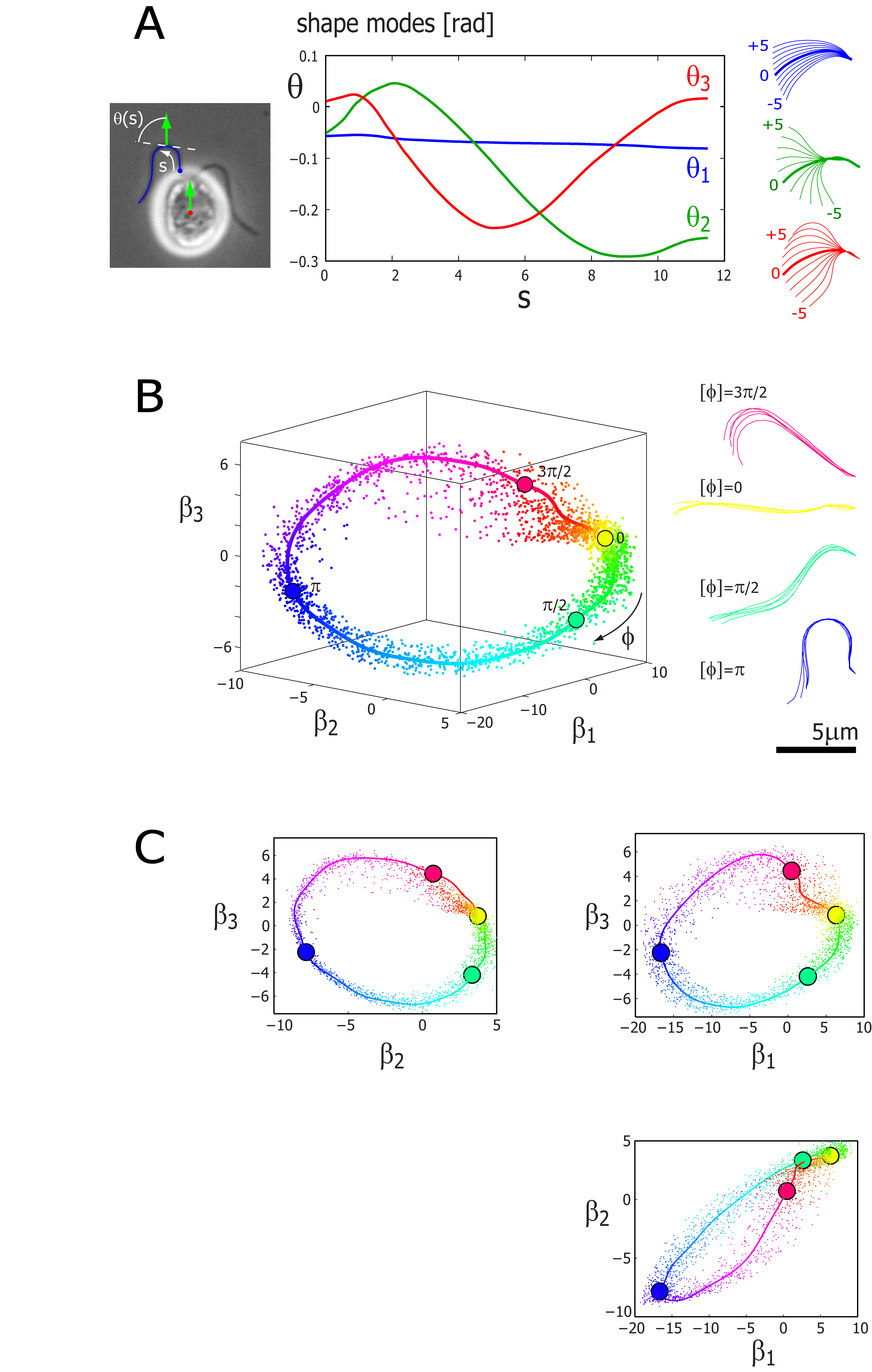}
\caption{
\label{fig_shape_circle}
We represent a single flagellar shape by $n=3$ shape coefficients as a point in an abstract shape space
that is spanned by $3$ principal shape modes.
\textbf{A.}
The principal shape modes were determined 
by employing a kernel principal component analysis (PCA) to the tangent angle representation $\theta(s)$ 
of smoothed flagellar shapes that were tracked from the left flagellum of cell no.~2 
during one beat cycle of synchronized flagellar beating.
From the PCA, we obtained three dominant shape modes with respective tangent angle representations 
$\theta_1(s)$, $\theta_2(s)$, $\theta_3(s)$ as shown.
Together, these principal shape modes account for $97\%$ of the variance of this tangent angle data set.
For sake of illustration, exemplary flagellar shapes
corresponding to the superposition of the mean flagellar shape and just one shape mode 
with tangent angle $\overline{\theta}(s)+\beta_i\theta_i(s)$, $i=1,2,3$ 
are shown to the right ($-5\le\beta_i\le 5$).
\textbf{B.}
Each tracked flagellar shape from one flagellum can be represented by a single point
in an abstract shape space that is spanned by the three principal shape modes.
More specifically, the coordinates $(\beta_1,\beta_2,\beta_3)$ of this point
are obtained by approximating the tracked flagellar shape by a superposition 
of a previously computed mean flagellar shape and the three principal shape modes, see equation~\ref{eq_fit_beta}.
The set of flagellar shapes from an entire experimental movie thus corresponds to a point cloud.
This point cloud scatters around a closed curve (solid line), which reflects the periodic nature of the flagellar beat.
This closed curve has been obtained by a simple fit to the point cloud of flagellar shapes
and can be considered as a limit cycle of flagellar beating.
Deviations from this limit cycle measure the variability of the flagellar beat.
We can use this representation to define a distinct flagellar phase angle $\varphi$ (modulo $2\pi$)
for each tracked flagellar shape as indicated by the color code
by mapping each flagellar shape onto the limit cycle.
A time-series of flagellar shapes thus yields a time-series of the flagellar phase angle $\varphi(t)$.
As an illustration of this assignment, 
superpositions of flagellar shapes are shown to the right, 
each of which correspond to flagellar shapes that were assigned the same flagellar phase modulo $2\pi$.
\textbf{C.}
Two-dimensional projections corresponding to the three-dimensional shape space representation in panel B.
}
\end{figure}

We employ a non-linear dimension reduction technique to represent tracked flagellar shapes
as points in a low-dimensional abstract shape space.
In a first step, smoothed tracked flagellar shapes corresponding to one cycle of synchronized 
flagellar beating (shown in figure \ref{fig_dof}A) were used to define the basis of the shape space.
Flagellar shapes can be conveniently represented
with respect to the material frame of the cell using a tangent angle representation \cite{Riedel:2007,Friedrich:2010}.
In terms of this tangent angle $\theta(s)$,
the $x(s)$ and $y(s)$ coordinates of the flagellar midline 
as functions of arclength $s$ along the flagellum can be expressed as
\begin{equation}
x(s)=x(0)+\int_0^s d\xi\,\cos[\alpha+\theta(\xi)]\text{ and }
y(s)=y(0)+\int_0^s d\xi\,\sin[\alpha+\theta(\xi)].
\end{equation}
Here, $\alpha$ is the orientation angle of the long axis of the cell body (figure \ref{fig_dof}A),
which implies that $\theta(\xi)$ characterizes flagellar shapes with respect to a material frame of the cell body.
By averaging the tangent angle profiles $\theta(s,t)$ over a full beat cycle, we define a time-averaged flagellar shape
characterized by a tangent angle $\overline{\theta}(s)$.
To characterize variations from this mean flagellar shape, 
we employed a kernel principal component analysis (PCA) \cite{Scholkopf:kernel}.
The kernel used to compute the Gram matrix $\mathrm{D}$ for the kernel PCA 
must account for the $2\pi$-periodicity of the tangent angle data
and was taken as $D_{ij}=\int_0^L ds\, \cos[\theta(s,t_i)-\theta(s,t_j)]$.
The first three shape eigenmodes account for 97\% of the spectrum of $\mathrm{D}$
and are shown in figure \ref{fig_shape_circle}A.
The relative contributions to the spectrum read $67\%$ (first mode), $18\%$ (second mode), $12\%$ (third mode).
While the first mode $\theta_1(s)$ (blue) describes nearly uniform bending of the flagellum,
the second mode $\theta_2(s)$ (green) and the third mode $\theta_3(s)$ (red) together
comprise the components of a traveling bending wave.

Next, any flagellar shape can be projected onto the shape space spanned by these three shape modes:
Given a flagellar midline with coordinates $x(s)$ and $y(s)$,
we seek the optimal approximating shape with coordinates $\hat{x}(s)$, $\hat{y}(s)$ 
whose tangent angle $\hat{\theta}(s)$ is a linear combination of the fundamental shape modes 
\begin{equation}
\label{eq_fit_beta}
\hat{\theta}(s) = \overline{\theta}(s) + \beta_1 \theta_1(s) + \beta_2 \theta_2(s) + \beta_3 \theta_3(s).
\end{equation}
The coefficients $\beta_1$, $\beta_2$, $\beta_3$ are obtained by a non-linear fit that minimizes
the squared Euclidean distance $\int_0^{L'} ds |x(s)-\hat{x}(s)|^2+|y(s)-\hat{y}(s)|^2$.
This procedure is robust and works even if flagellar shapes could only be tracked partially
with tracked length $L'$ shorter than the total flagellar length $L$.
Note that for non-smoothed flagellar shapes, 
the tangent angle representations can be noisy and are thus less suitable for fitting as compared to $x,y$ coordinates.

A time sequence of tracked flagellar shapes thus results in a point cloud 
in the shape space parametrized by the shape mode coefficients $\beta_1$, $\beta_2$, $\beta_3$.
We fitted a closed curve to the torus-like point cloud, see the solid line in figure \ref{fig_shape_circle}B.
This closed curve represents a limit cycle of periodic flagellar beating.
Each tracked flagellar shape can be assigned the `closest' point on this limit cycle,
\textit{i.e.}\ the point for which the corresponding flagellar shape has minimal Euclidean distance.
By choosing a phase angle parametrization for the limit cycle, 
the phase angle of each flagellar shape is determined modulo $2\pi$.
A time-series of flagellar shapes thus yields a time-series of the flagellar phase angle $\varphi(t)$.
The phase angle parametrization of the limit cycle had been chosen 
such that the flagellar phase angle $\varphi$ and its time derivative are not correlated.
Finally, the zero point $\varphi=0$ was chosen such that the corresponding flagellar shape was nearly straight and perpendicular
to the long cell axis.

\subsection{Computation of hydrodynamic friction forces}

For our hydrodynamic computations, 
we represented a \textit{Chlamydomonas} cell by an ensemble of $N=300$ equally sized spheres of radius $a=0.25\,\micron$.
The cell body was chosen spheroidal and is represented by $272$ spheres 
that are arranged in a symmetric fashion to retain mirror symmetries.
Each flagellum is represented by a chain of $14$ spheres that are aligned along a flagellar midline with equidistant spacing.
The shapes of the flagellar midlines depend on respective phase angles $\varphi_L$ and $\varphi_R$ for the left and right flagellum.
These flagellar shapes were taken from experiment for one full period of synchronized beating and are shown in figure \ref{fig_dof}A.
We assume that the $272$ spheres constituting the cell body move as a rigid sphere cluster.
Each of the flagellar spheres represents a cluster with just one sphere, 
which results in a total of $n=2\cdot 14+1=29$ sphere clusters.
We then computed the $6n\times 6n$ grand hydrodynamic friction matrix $\mathrm{G}$ for this ensemble of $n$ spheres clusters
using a freely available hydrodynamic library based on a Cartesian multipole expansion technique \cite{Hinsen:1995}. 
Recall that the grand hydrodynamic friction matrix $\mathrm{G}$ relates the forces and torques 
exerted by the $6n$ sphere clusters to their translational and rotational velocities \cite{Happel:hydro}
\begin{equation}
\mathrm{P}_0 = \mathrm{G} \cdot \dot{\mathrm{q}}_0.
\end{equation}
Here, $\dot{\mathrm{q}}_0$ denotes a $6n$-vector that combines the 
translational and rotational velocity components of the $n$ sphere clusters,
\begin{equation}
\dot{\mathrm{q}}_0=(v_{1x},v_{1y},v_{1z},\omega_{1x},\omega_{1y},\omega_{1z},\ldots,\omega_{nz}),
\end{equation}
while the $6n$-vector $\mathrm{P}_0$ combines the components of the resultant hydrodynamic friction forces and torques,
\begin{equation}
\mathrm{P}_0=(F_{1x},F_{1y},F_{1z},T'_{1x},T'_{1y},T'_{1z},\ldots,T'_{nz}).
\end{equation}
(Primed torques represent torques with respect to the center of the respective sphere cluster.)
Figure \ref{fig_hydro}C,D in the main text shows a sub-matrix of the grand friction matrix, 
which was defined as $G_{ij,yy}=G_{6i-4,6j-4}$, $i,j=1,\ldots,n$.
In this figure, it was assumed that the long cell body axis is aligned with the $y$-axis of the laboratory frame, 
\textit{i.e.} $\alpha=0$, which implies that the sub-matrix relates
motion in the direction of the long cell axis and the hydrodynamic force components projected on this axis.

For our hydrodynamic computations, the multipole expansion order was chosen as $3$.
An estimate for the accuracy of our computation could be obtained by increasing the expansion order parameter,
which changed the computed friction coefficients by less than $1\,\%$.
Initial tests confirmed that the friction matrix of only the cell body gave practically the same result as the analytic solution for the enveloping spheroid \cite{Perrin:1934}; similarly the computed friction matrix of only a single flagellum matched the prediction of resistive force theory \cite{Gray:1955b} assuming a flagellar radius equal to the sphere radius. 
Note that the precise value of the flagellar radius is expected to affect hydrodynamic friction coefficients 
only as a logarithmic correction \cite{Brennen:1977}.

Below, we consider an extension of the theoretical description given in the main text
that additionally considers the possibility of an elastically anchored flagellar base,
which allows for pivoting of the flagellar basal apparatus, see figure~\ref{fig_sync_pivo}.
In this case, the flagellar midlines were rotated by an angle $\psi$.

A set of 2400 pre-computed configurations was then used to construct a spline-based lookup-table 
of the (reduced) hydrodynamic friction matrix as a function of the degrees of freedom $\varphi_L$, $\varphi_R$ and $\psi$.
The interpolation error was confirmed to be on the order of $1\,\%$ or less.
This lookup-table was then used for the numerical integration of the 
(stiff) equations of motion \ref{eq_motion} and \ref{eq_motion_pivo}.

\subsection{Generalized hydrodynamic friction forces}

\begin{figure}[b]
\includegraphics[width=8.5cm]{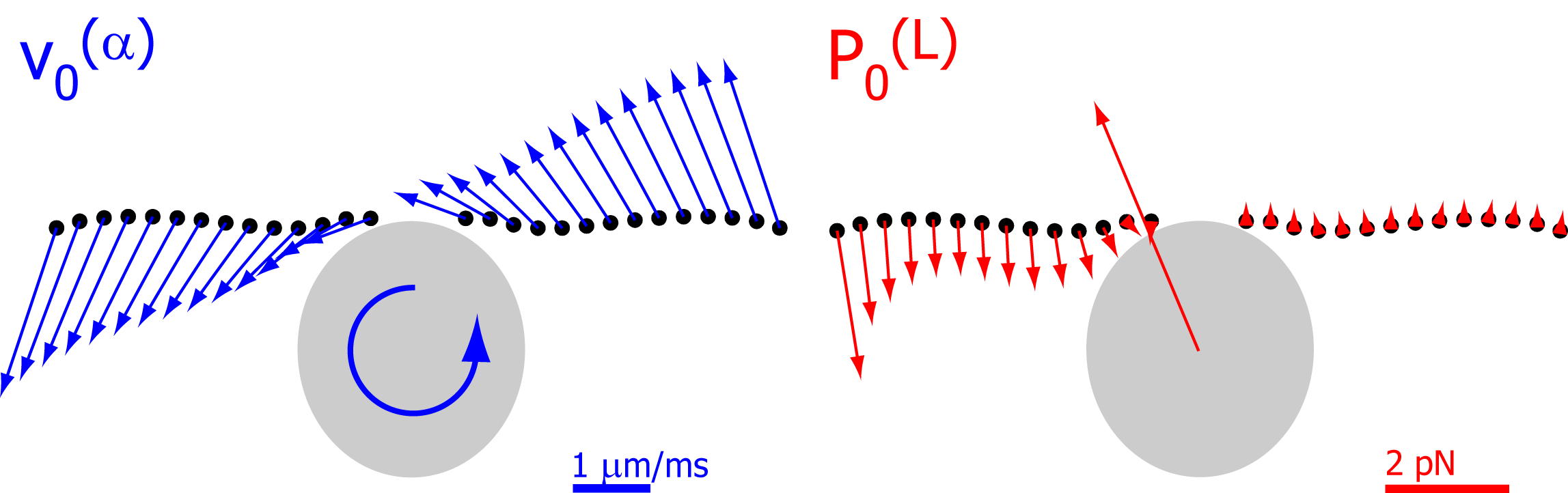}
\caption{
\label{fig_g34}
Coupling of cell body yawing and flagellar beating.
\textbf{(left)}
Translational velocities of the flagellar spheres used in our hydrodynamic computation
associated with a pure yawing motion of the cell body with rate $\dot{\alpha}$.
\textbf{(right)}
Hydrodynamic friction forces exerted by the flagellar spheres (as well as by the cell body),
if the left flagellum advances along its beat cycle with rate $\dot{\varphi}_L$.
The generalized hydrodynamic friction coefficient $\Gamma_{\alpha L}$ 
that couples cell body yawing and beating of the left flagellum
can be computed as a scalar product 
between the velocity profile resulting from yawing
and the force profile resulting from flagellar beating
and is found to be non-zero.
Parameters: $\dot{\varphi}_L=\omega_0$, $\dot{\alpha}=0.2\omega_0$, $2\pi/\omega_0=30\,\mathrm{ms}$.
}
\end{figure}

We employ the framework of Lagrangian mechanics of dissipative systems \cite{Goldstein:mechanics}
to define generalized hydrodynamic friction forces and derive an equation of motion 
for the effective degrees of freedom in our theoretical description of \textit{Chlamydomonas} swimming and synchronization.
The $6n$ degrees of freedom $\mathrm{q}_0$ for the $n$ sphere clusters
used in our hydrodynamic computations are enslaved by the 5 effective degrees of freedom in our coarse-grained theory, 
see figure \ref{fig_dof}. 
Below, one more degree of freedom $\psi$ is introduced to characterize pivoting of 
an elastically anchored flagellar basal apparatus.
We thus have
\begin{equation}
\mathrm{q}_0 = \mathrm{q}_0(\mathrm{q}),
\end{equation}
where we introduced the $6$-component vector 
$\q=(x,y,\alpha,\varphi_L,\varphi_R,\psi)$ that comprises the $6$ effective degrees of freedom.
The reduced $6\times 6$ hydrodynamic friction matrix $\Gamma$ for these 
$6$ effective degrees of freedom 
can be computed from the grand hydrodynamic friction matrix $\mathrm{G}$ as
\begin{equation}
\label{eq_LGL}
\Gamma=\mathrm{L}^T \cdot\mathrm{G}\cdot \mathrm{L}
\end{equation}
with a $6n\times 6$ transformation matrix $\mathrm{L}$ given by \cite{Friedrich:2012c}
\begin{equation}
L_{ij}=\partial{\dot{q}_{0,i}}/\partial{\dot{q}_j}.
\end{equation}
The rate of hydrodynamic dissipation can now be equivalently written as
a quadratic function of either $\dot{\mathrm{q}}_0$ or $\dot{\mathrm{q}}$
\begin{equation}
\mathcal{R} = 
\dot{\mathrm{q}}_0^T\cdot\mathrm{G}\cdot\dot{\mathrm{q}}_0 = 
\dot{\mathrm{q}}^T\cdot\Gamma\cdot\dot{\mathrm{q}} .
\end{equation}
The generalized hydrodynamic friction coefficients $\Gamma_{ij}$ are
depicted in figure \ref{fig_gh}.
In this context, generalized hydrodynamic friction forces can be defined as
\begin{equation}
P_j = \Gamma_{jx}\dot{x} + \Gamma_{jy}\dot{y} + \Gamma_{j\alpha}\dot{\alpha}
+ \Gamma_{jL}\dot{\varphi}_L + \Gamma_{jR}\dot{\varphi}_R + \Gamma_{j\psi}\dot{\psi},
\quad
j=x,y,\alpha,L,R,\psi.
\end{equation}
Interestingly, the generalized hydrodynamic friction force conjugated
to one degree of freedom depends also on the rates of the change of the other degrees of freedom,
which implies a coupling between the various degrees of freedom.
This fact is illustrated by figure~\ref{fig_g34}.
Panel A depicts 
the translational velocities of the flagellar spheres
caused by pure yawing of the cell body with rate $\dot{\alpha}$.
This motion is characterized by a $6n$-vector of velocity components,
$\dot{\mathrm{q}}_0^{(\alpha)}=\mathrm{L}\cdot(0,0,\dot{\alpha},0,0,0)^T$.
Similarly, the beating of the left flagellum induces 
hydrodynamic friction forces as shown in figure~\ref{fig_g34}B.
The resultant force (and torque) components are combined in the $6n$-vector
$\mathrm{P}_0^{(L)}=\mathrm{G}\cdot\mathrm{L}\cdot(0,0,0,\dot{\varphi}_L,0,0)^T$.
Figure~\ref{fig_g34} indicates that the scalar product 
$\dot{\mathrm{q}}_0^{(\alpha)}\cdot\mathrm{P}_0^{(L)}=\dot{\alpha}\Gamma_{\alpha L}\dot{\varphi}_L$ does not vanish,
which implies a non-zero friction coefficient $\Gamma_{\alpha L}$ 
and thus a coupling between cell body yawing and flagellar beating.

In our theoretical description,
the phase dynamics of the left flagellum, say, is
governed by a balance of the generalized hydrodynamic friction force $P_L$ and
an active driving force $Q_L$; similarly, $Q_R=P_L$ for the right flagellum.
In the case of free swimming,
force and torque balance imply $P_x=P_y=0$ and $P_\alpha=0$.
Together with an equation for $P_\psi$, these equation
allow to self-consistently solve for the rate of change $\dot{\mathrm{q}}$
of the $6$ degrees of freedom.
If one degree of freedom were constrained, $q_j=0$,
the corresponding force equation becomes void, 
since a constraining force $Q_j$ equal to $P_j$ then balances 
the generalized hydrodynamic friction force $P_j$ associated with this degree of freedom.

In general, the active driving forces $Q_L$ and $Q_R$ will depend on the flagellar phase.
This phase-dependence is fully determined by the requirement that the flagellar
phase speeds should be constant, $\dot{\varphi}_j=\omega_0$, 
in the case of synchronized flagellar beating with $\delta=0$.
Here, $\omega_0$ denotes the angular frequency of synchronized flagellar beating.
Explicitly, we find
\begin{equation}
Q_L(\varphi_L) = \omega_0 [ \Gamma_{LL}(\varphi_L,\varphi_L) +
\Gamma_{LR}(\varphi_L,\varphi_L) -
2\Gamma^2_{Ly}(\varphi_L,\varphi_L)/\Gamma_{yy}(\varphi_L,\varphi_L)  ].
\end{equation}
An analogous expression holds for $Q_R(\varphi_R)$.
Note that the generalized active driving forces are conjugate to an angle,
and therefore have the physical unit 
$\mathrm{pN}\,\mu\mathrm{m}/\mathrm{rad}$.
These phase-dependent active driving forces can be written as potential forces
$Q_j=-\partial U/\partial\varphi_j$, $j=L,R$, where the potential $U$ reads
\begin{equation}
U=-\int_{-\infty}^{\varphi_L} d\varphi_L\, Q_L(\varphi_L) - \int_{-\infty}^{\varphi_R}d\varphi_R \, Q_R(\varphi_R).
\end{equation}
The potential $U$ continuously decreases with time,
indicating the depletion of an internal energy store and the dissipation of energy into the fluid
during flagellar swimming.
The rate of hydrodynamic dissipation equals the rate at which potential energy is dissipated
\begin{equation}
\mathcal{R}=-\dot{U}=Q_L\dot{\varphi}_L+Q_R\dot{\varphi}_R.
\end{equation}

\subsection{Analytic expression for the flagellar synchronization strength}

We present details on the derivation of equations \ref{eq_lyaw} and \ref{eq_lelastic} 
for the synchronization strength $\lambda$
in the case of the reduced equations of motion \ref{eq_yaw1}-\ref{eq_yaw3}.
We assume equal intrinsic beat frequencies, $\omega_L=\omega_R=\omega_0$
and a small initial phase difference, $0<\delta(0)\ll 1$.
To leading order in $\delta$, we find relations
that link the rotation rate $\dot{\alpha}$ and the rate $\dot{\delta}$ at which the phase difference changes,
\begin{align}
\label{eq_alphadelta1}
\kyaw\alpha+\rho(\varphi,\varphi)\dot{\alpha} &= -d[\nu(\varphi)\delta]/dt \\
\label{eq_alphadelta2}
\dot{\delta} &= -2\mu(\varphi)\dot{\alpha}.
\end{align}
Here $\varphi\approx\omega_0 t$ denotes the mean flagellar phase. 
The first equation describes how flagellar asynchrony causes a yawing motion of the cell body,
while the second equation describes how this yawing motion then changes the flagellar phase difference.
In the absence of any elastic constraint for yawing, $k=0$, 
we can solve for $\dot{\delta}$
\begin{equation}
\label{eq_dyaw}
(\rho-2\mu\nu)\dot{\delta} = 2\mu \nu' \omega_0 \delta.
\end{equation}
Now, equation \ref{eq_lyaw} follows from equation \ref{eq_dyaw}
using $\lambda = - \int_0^T dt\,\dot{\delta}/\delta$
and a variable transformation $\varphi(t)=\omega_0 t+\mathcal{O}(\delta)$.

In the case of a very stiff elastic constraint with $k\gg\rho\omega_0$,
we make use of the fact that variations of the phase difference $\delta$ during one beat cycle 
will be small compared to its mean value $\delta_0=\langle\delta\rangle$.
As a consequence, equation \ref{eq_alphadelta1} can be approximated as
$\kyaw\alpha = - \nu'\omega_0\delta_0$.
Using this approximation and equation \ref{eq_alphadelta2},
equation \ref{eq_lelastic} follows.

\subsection{Comparison of experiment and theory}

\begin{figure}
\includegraphics[width=8.5cm]{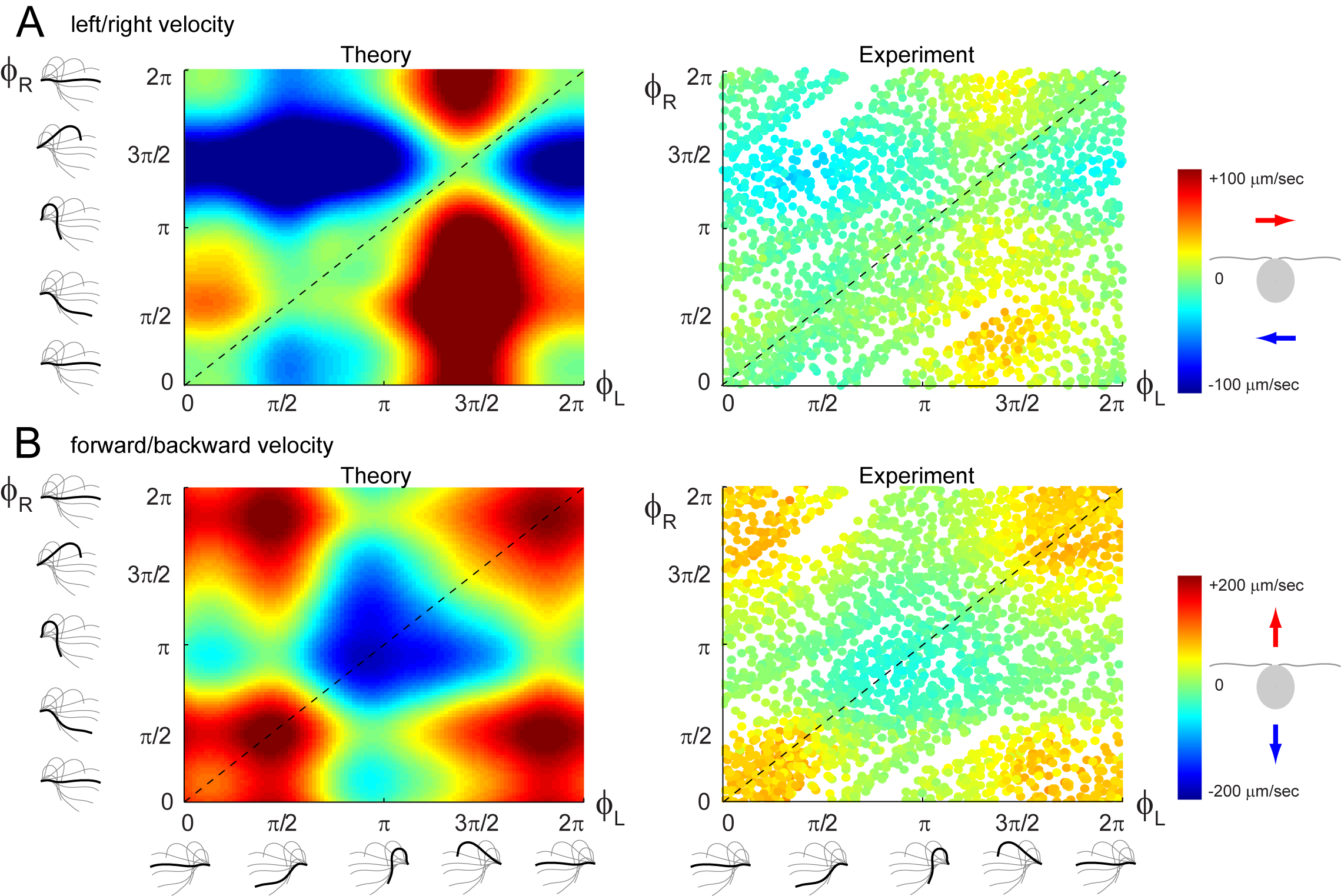}
\caption{
\label{fig_v1_v2}
\textbf{A.} 
Instantaneous swimming velocity in the direction perpendicular to the long cell axis
as a function of the flagellar phase angles $\varphi_L$ and $\varphi_R$.
For synchronized flagellar beating (dashed line), 
this velocity vanishes in our theory for symmetry reasons (green).
If the two flagella are out of synchrony, however, significant sidewards motion of the cell is observed,
both in theory and experiment.
Note that wall effects present in the experiments, but not considered in the computations,
reduce translational velocities.
\textbf{B.}
Instantaneous swimming velocity in the direction of the long cell axis,
again as a function of the flagellar phase angles.
}
\end{figure}

\begin{figure}
\includegraphics[width=8.5cm]{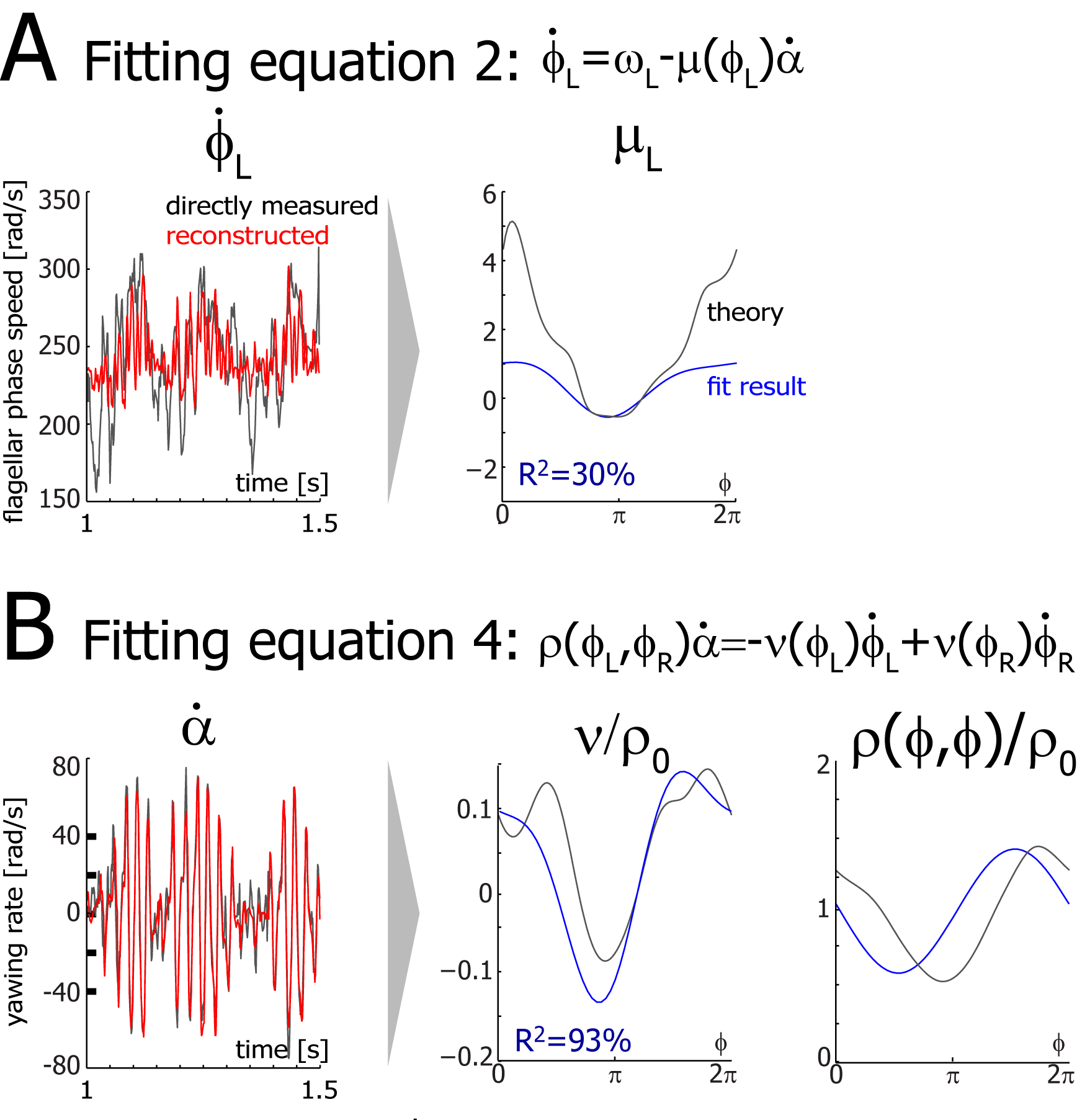}
\caption{
\label{fig_one_fit}
The reduced equations of motion \ref{eq_yaw1}-\ref{eq_yaw3}
were fitted to experimental time series data for the yawing rate $\dot{\alpha}$ of the cell,
as well as the flagellar phase speeds $\dot{\varphi}_L$ and $\dot{\varphi}_R$.
This provided experimental estimates for the 
phase-dependent coupling functions $\mu$, $\nu$, and $\rho$.
Specifically, 
we represented each coupling function as a truncated Fourier series and
determined the unknown Fourier coefficients by a linear regression using eqs.~\ref{eq_yaw1}-\ref{eq_yaw3}.
\textbf{A.} 
Linear regression of equation~\ref{eq_yaw1}.
Shown to the left in black is the instantaneous flagellar phase speed of the left flagellum
$\dot{\varphi}_L$ (smoothed with a span of $15\,\mathrm{ms}$).
Shown in red is a reconstructed phase speed 
$\omega_L-\mu_L(\varphi_L)\dot{\alpha}$
that depends on the instantaneous cell body yawing rate $\dot{\alpha}$,
as well as the intrinsic flagellar frequency $\omega_L$ 
and phase-dependent coupling function $\mu_L$ for the left flagellum, which were obtained by the fit.
The coefficient of determination was $R^2=30\%$.
The estimate for $\mu_L$ obtained from this fit is shown to the right (blue),
together with a theoretical prediction (black), see also figure~\ref{fig_yaw}A.
\textbf{B.}
Linear regression of equation~\ref{eq_yaw3}.
Shown to the left is the measured instantaneous yawing rate $\dot{\alpha}$ of the cell body (black),
and a yawing rate reconstructed from the flagellar phase dynamics,
$[-\nu(\varphi_L)\dot{\varphi}_L+\nu(\varphi_R)\dot{\varphi}_R]/\rho(\varphi_L,\varphi_R)$ (red).
The coefficient of determination was $R^2=93\%$.
From this fit, we obtain an experimental estimate for the coupling functions
$\nu(\varphi)$ and $\rho(\varphi_L,\varphi_R)$ (blue);
theoretical predictions are shown in black.
}
\end{figure}

\begin{figure}
\includegraphics[width=17cm]{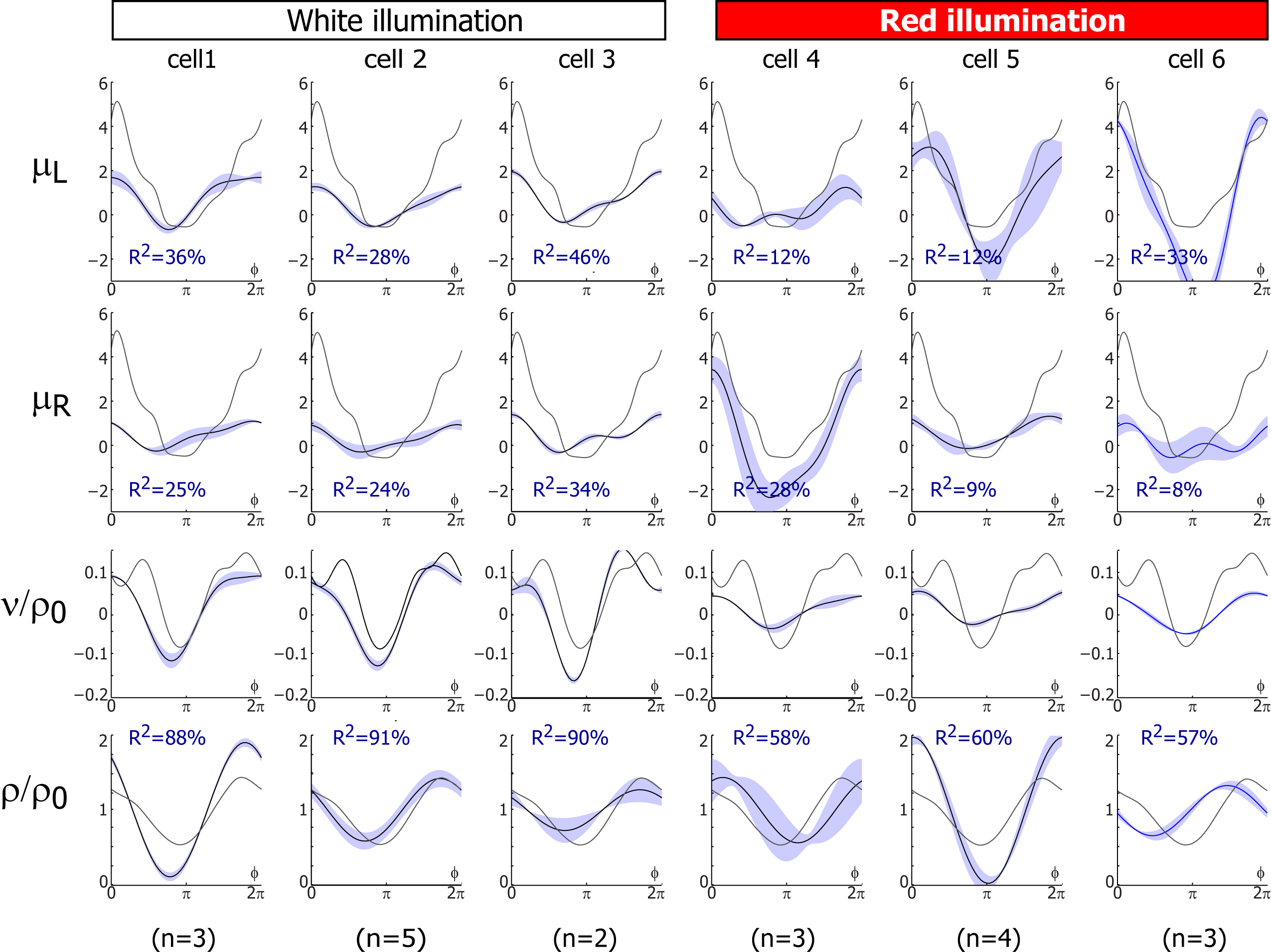}
\caption{
\label{fig_fits}
Experimental fits for the coupling functions $\mu$, $\nu$, and $\rho$ introduced 
in eqns.~\ref{eq_yaw1}-\ref{eq_yaw3} (blue curves, shaded regions indicate mean$\pm$s.e.)
and theoretical predictions (black).
The coupling functions $\mu$, $\nu$, and $\rho$ relate flagellar beating and cell body yawing.
Fitting results are shown for 6 different cells illuminated by either white or red light as indicated.
For each cell, we employed $n$ fits using $n$ non-overlapping time series of duration $0.4-0.5\,\mathrm{s}$ with $n$ as indicated.
The blue curves represent the average of the fitted coupling functions for the $n$ fits;
the averaged coefficient of determination $R^2$ is stated.
}
\end{figure}

We can compare instantaneous swimming velocities predicted by our hydrodynamic computation
with experimental measurements and find favorable agreement, see figures \ref{fig_swim} and \ref{fig_v1_v2}.
Note that wall effects present in our experiments, but not accounted for by our hydrodynamic computations,
are expected to reduce translational velocities (but less so rotational velocities) \cite{Bayly:2011}.
The hydrodynamic computations are based on a fixed flagellar beat pattern parametrized by a flagellar phase angle,
which was obtained experimentally for one beat cycle with synchronized beating (shown in figure \ref{fig_dof}A).
The good agreement between theoretical predictions and experimental measurements for the instantaneous swimming velocities
further validate our reductionist description of the flagellar shape dynamics by just a single phase variable for each flagellum.
Next, we tested the applicability of the reduced equations of motion eqs.~\ref{eq_yaw1}-\ref{eq_yaw3} in the experimental situation.
For this aim, we reconstructed the coupling functions $\mu(\varphi)$, $\nu(\varphi)$ and $\rho(\varphi)$
from experimental time series data for $\dot{\alpha}$, $\dot{\varphi}_L$, and $\dot{\varphi}_R$.
The coupling functions were represented by truncated Fourier series and 
the unknown Fourier coefficients determined by a linear regression of equation~\ref{eq_yaw1}, \ref{eq_yaw2}, or \ref{eq_yaw3},
respectively, see figure \ref{fig_one_fit}.
Repeating this fitting procedure for data from 6 different cells gave consistent results, see \ref{fig_fits}.
Moreover, the phase-dependence of the fitted coupling functions 
agrees qualitatively with our theoretical predictions.
Note that our simple theory does not involve any adjustable parameters.

\subsection{An elastically anchored flagellar basal apparatus}
\label{sec_pivo}

\begin{figure}
\includegraphics[width=8.5cm]{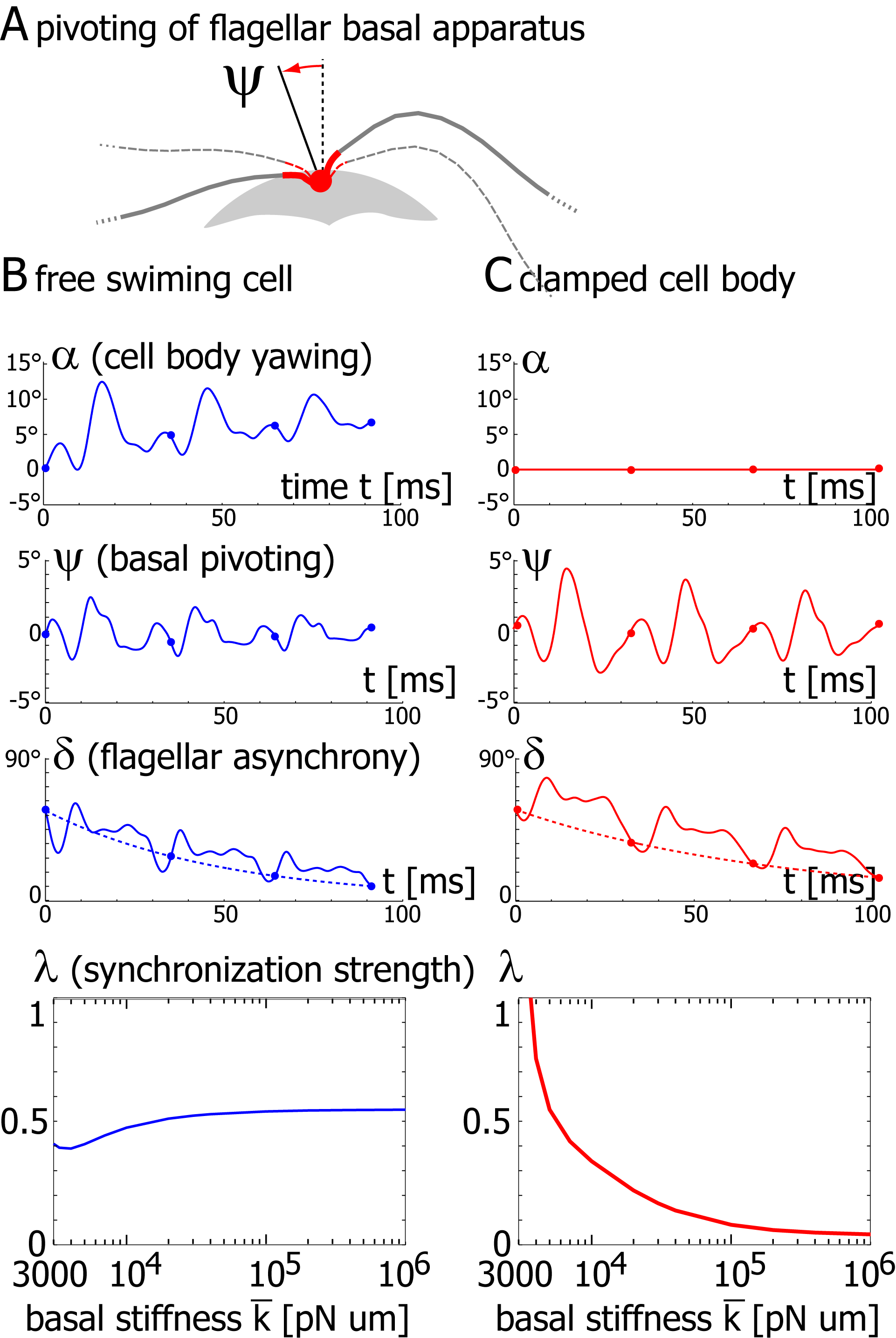}
\caption{
\label{fig_sync_pivo}
Theory of flagellar synchronization by basal pivoting.
\textbf{A.} 
We consider the possibility of an elastically anchored flagellar basal apparatus (red),
which allows for pivoting of the basal apparatus (solid lines) by an angle $\psi$ from its symmetric
reference configuration (dashed).
\textbf{B.}
For a free swimming cell,
the equation of motion \ref{eq_motion_pivo} predicts both 
a yawing motion of the cell characterized by $\alpha(t)$
and a pivoting motion of the flagellar base characterized by $\psi(t)$,
if the two flagella are initially out of synchrony.
The flagellar phase differences $\delta$ is found to decrease with time (solid line),
approximately following an exponential decay (dotted line).
This implies that the in-phase synchronized state is stable with respect to perturbations.
Each completion of a full beat cycle of the left flagellum is marked by a dot.
The synchronization behavior in the case of an elastically anchored flagellar basal apparatus 
is nearly identical to the case of a stiff anchorage as shown in the main text in figure \ref{fig_sync}.
The lowest panel shows 
typical amplitudes of basal pivoting ($\delta\psi$, solid line) and cell body yawing ($\delta\alpha$, dashed) 
as a function of basal stiffness $\kpivo$.
Amplitudes were determined as half the range of variation during one beat cycle
for an initial phase difference of $\delta(0)=\pi/2$.
\textbf{C.}
For a clamped cell that can neither translate nor rotate,
the flagellar apparatus can still pivot and will do so if the two flagella are initially out of phase.
As in the case of a free-swimming cell, 
the flagellar phase difference $\delta$ decays with time, indicating stable synchronization.
In the case of a clamped cell, the synchronization strength $\lambda$ strongly depends
on the stiffness $\kpivo$ of the elastic anchorage of the basal flagellar apparatus,
which sets the amplitude of basal pivoting.
Parameters: $2\pi/\omega_0=30\,\mathrm{ms}$, $\kpivo=10^4\,\mathrm{pN}\,\mu\mathrm{m/rad}$.
}
\end{figure}

In the main text, we had assumed for simplicity that the flagellar base is rigidly anchored to the cell body.
While the proximal segments of the two flagella are tightly mechanical coupled with each other by so-called 
striated fibers to form the flagellar basal apparatus, 
the flagellar basal apparatus itself is only connected to an array of $16$ long microtubules spanning the cell \cite{Ringo:1967}.
We now consider the possibility that this anchorage allows for some pivoting of the flagellar basal apparatus as a whole 
by an angle $\psi$, see figure \ref{fig_sync_pivo}A.
In addition to the five degrees of freedom of \textit{Chlamydomonas} beating and swimming considered in the main text
(see figure \ref{fig_dof}), we now include this pivot angle $\psi$ as a $6^\mathrm{th}$ degree of freedom.
The rate of hydrodynamic dissipation is now given by 
$\mathcal{R}=\dot{x} P_x + \dot{y} P_y + \dot{\alpha} P_\alpha + \dot{\varphi}_L P_L+\dot{\varphi}_R P_R + \dot{\psi} P_\psi$,
with $P_\psi$ being the generalized hydrodynamic friction force conjugate to the pivot angle $\psi$.
Assuming Hookean behavior for the elastic basal anchorage with rotational pivoting stiffness $\kpivo$, 
we readily arrive at an equation of motion that reads in the case of free swimming
\begin{equation}
\label{eq_motion_pivo}
(\dot{x},\dot{y},\dot{\alpha},\dot{\varphi}_L,\dot{\varphi}_R,\dot{\psi})^T =
\Gamma^{-1}(0,0,0,Q_L,Q_R,-\kpivo\psi)^T.
\end{equation}

Figure \ref{fig_sync_pivo}B shows flagellar synchronization for a free swimming cell with elastically anchored flagellar base:
Although some basal pivoting occurs as a result of flagellar asynchrony,
the swimming and synchronization behavior is very similar to the case of a rigidly anchored flagellar base
as shown in figure \ref{fig_sync}A.
For a cell that can neither translate nor yaw, however, the situation is different, see figure \ref{fig_sync_pivo}C.
We find strong flagellar synchronization provided the elastic stiffness $\kpivo$ is not too large.
Flagellar synchronization by basal pivoting is thus effective also for a fully clamped cell.
In contrast, for a rigidly anchored flagellar base, the synchronization strength $\lambda$ would be relatively weak
in this case, being due only to direct hydrodynamic interactions between the two flagella.

\begin{figure}[b]
\includegraphics[width=8.5cm]{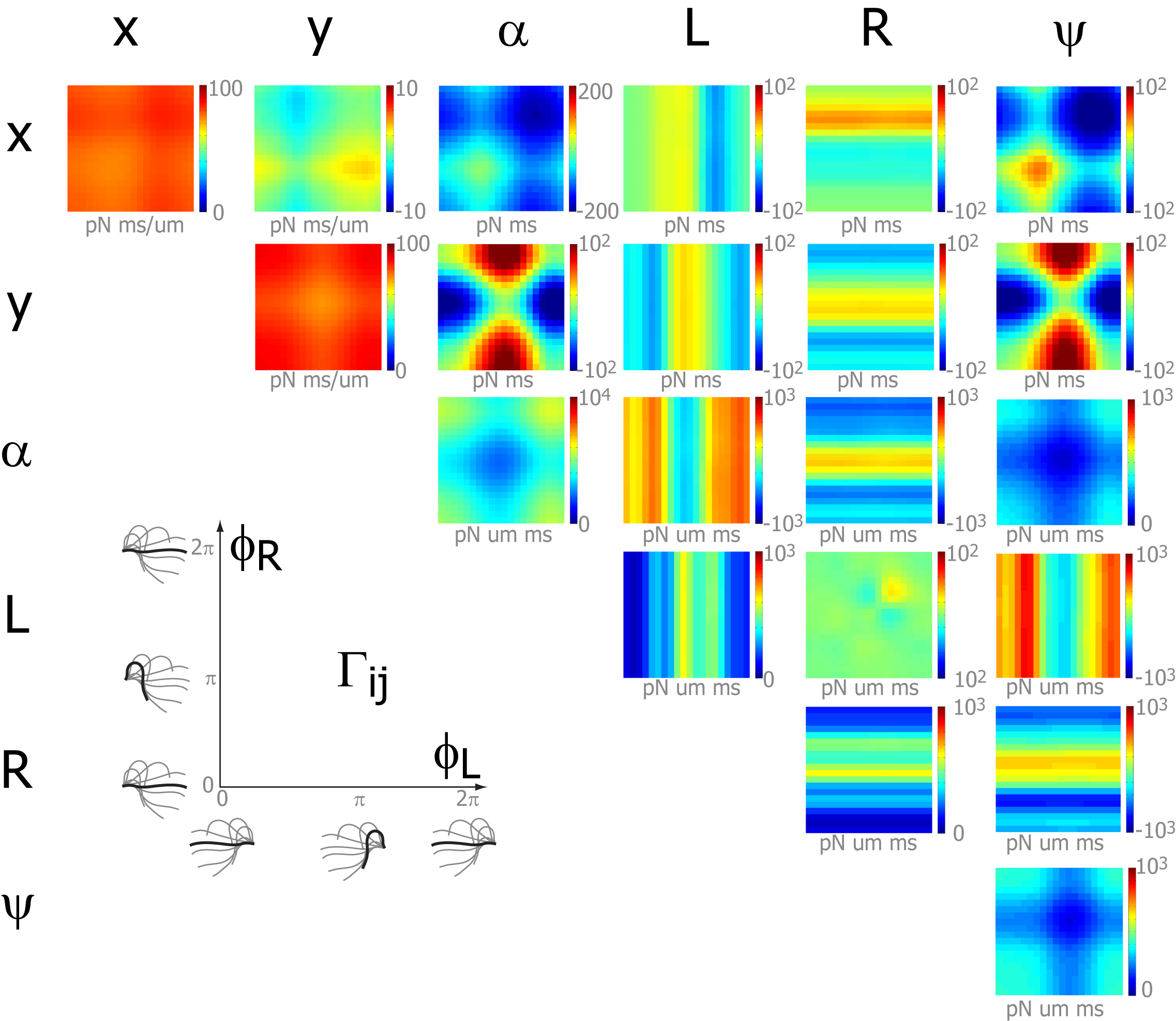}
\caption{
\label{fig_gh}
Generalized hydrodynamic friction matrix $\Gamma_{ij}$ 
associated with the effective degrees of freedom $x$, $y$, $\alpha$, $\varphi_L$, $\varphi_R$, $\psi$.
This generalized friction matrix 
determines the generalized hydrodynamic friction forces $P_i$ conjugate
to the degrees of freedom $\mathrm{q}=(x,y,\alpha,\varphi_L,\varphi_R,\psi)$ as 
$P_i = \Gamma_{ij} \dot{q}_j$,
and is computed as a projection of the grand
hydrodynamic friction matrix, see equation~\ref{eq_LGL}.
Each friction coefficient $\Gamma_{ij}$ is a periodic function of the two phase angles $\varphi_L$ and $\varphi_R$,
$\Gamma_{i,j}=\Gamma_{i,j}(\varphi_L,\varphi_R)$
and is represented as a colorplot with axes as indicated.
Here, $\alpha$ is set to zero; different values of $\alpha$ would correspond to a simple rotation of the matrix shown.
By Onsager symmetry, $\Gamma_{ij}=\Gamma_{ji}$ \cite{Landau:hydro}. 
Several features are note-worthy:
The coefficient $\Gamma_{LR}$ characterizes hydrodynamic interactions between the two flagella, and is found to be small compared to e.g. $\Gamma_{LL}$.
The other coefficients $\Gamma_{Lj}=\Gamma_{jL}$, which set the friction force $P_L$ conjugate to $\varphi_L$, depend strongly on $\varphi_L$, but almost not on $\varphi_R$.
This is yet another manifestation of the fact that direct hydrodynamic interactions between the two flagella
are comparably weak.
Analogous statements hold for the coefficients $\Gamma_{Rj}$.
A counter-clockwise rotation of the cell, $\dot{\alpha}>0$, will increase the friction force $P_L$ during the effective stroke of the left flagellum
($\Gamma_{L\alpha}>0$), but decrease the corresponding the respective friction force $P_R$ for the right flagellum during its effective stroke
($\Gamma_{R\alpha}<0$).
Mirror-symmetry of the swimmer amounts to invariance of the friction matrix under the substitution
$(x,y,\alpha,\varphi_L,\varphi_R)\rightarrow (-x,y,-\alpha,\varphi_R,\varphi_L)$,
which implies a number of symmetry relations, 
\textit{e.g.} $\rho=\Gamma_{\alpha\alpha}$ must be symmetric in $\varphi_L$ and $\varphi_R$.
Finally, this rotational friction coefficient $\rho=\Gamma_{\alpha\alpha}$ depends on the flagellar phases in a more pronounced way
than the translational friction coefficients $\Gamma_{xx}$, $\Gamma_{yy}$. 
This is inline with the general fact that
rotational friction coefficients depend stronger (as $\sim l^3$) on the effective linear dimension $l$ of an object than translational friction coefficients ($\sim l$).
The coefficients $\Gamma_{j\alpha}$ and $\Gamma_{j\psi}$ associated with
yawing of the whole cell and pivoting of the flagellar apparatus, respectively,
show a similar dependence on the flagellar phases.
} 
\end{figure}

\begin{figure}
\includegraphics[width=8.5cm]{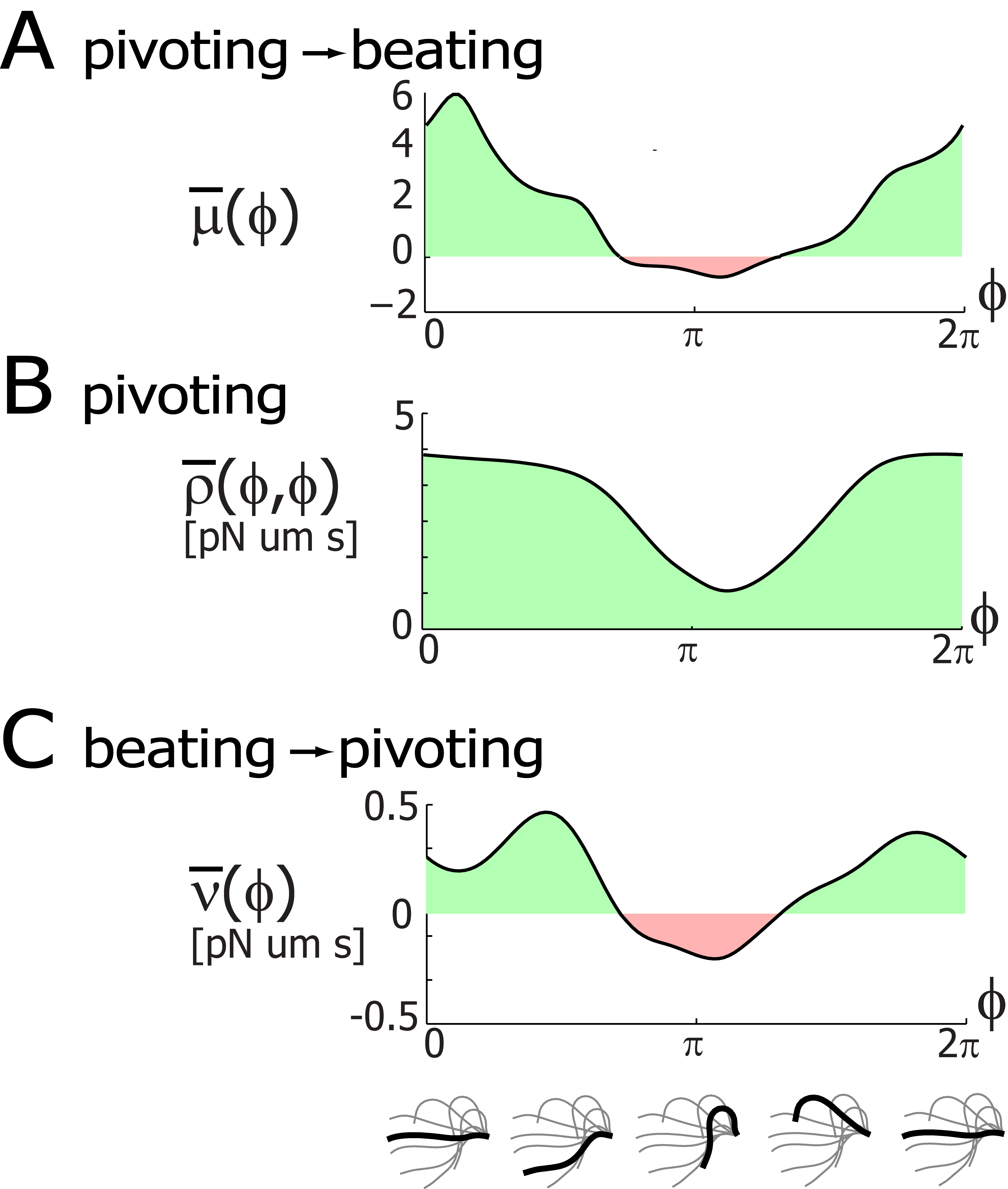}
\caption{
\label{fig_pivo}
Theoretical coupling functions for the case of a pivoting flagellar base.
\textbf{A.}
A pivoting motion of the flagellar base 
changes the hydrodynamic friction force associated with flagellar beating
and thereby speeds up or slows down the flagellar beat cycle in our theory.
This effect is quantified by a coupling function $\ol{\mu}$, see equation~\ref{eq_pivo1}.
\textbf{B.}
Hydrodynamic friction associated with pivoting of the flagellar base (and the attached flagella)
is characterized by a friction coefficient $\ol{\rho}$, see equation~\ref{eq_pivo3}.
This friction coefficient is maximal when the two flagella extend maximally from the cell body during their effective stroke.
\textbf{C.}
The beat of the left flagellum causes pivoting of the flagellar base.
This effect is quantified by a coupling function $\ol{\nu}$, see eqn.~\ref{eq_pivo3}.
}
\end{figure}

Flagellar synchronization by basal pivoting is conceptually 
very similar to synchronization by cell body yawing as discussed in the main text.
In the case of a fully clamped cell, we can approximate the synchronization dynamics 
by virtually the same generic equation of motion as eqs.~\ref{eq_yaw1}-\ref{eq_yaw3}, when we substitute $\psi$ for $\alpha$
\begin{align}
\label{eq_pivo1}
\dot{\varphi}_L &=\omega_0 -\ol{\mu}(\varphi_L)\dot{\psi} , \\
\label{eq_pivo2}
\dot{\varphi}_R &=\omega_0 +\ol{\mu}(\varphi_R)\dot{\psi} , \\
\label{eq_pivo3}
\kpivo\psi+\ol{\rho}(\varphi_L,\varphi_R)\dot{\psi} &= -\ol{\nu}(\varphi_L)\dot{\varphi}_L+\ol{\nu}(\varphi_R)\dot{\varphi}_R .
\end{align}
Here, the coupling functions $\ol{\mu}$, $\ol{\nu}$ and $\ol{\rho}$
play a similar role as the previously defined $\mu$, $\nu$ and $\rho$ for equation~\ref{eq_yaw1}-\ref{eq_yaw3}
and show a qualitatively similar dependence on the flagellar phase, see figure \ref{fig_pivo}.
To derive equations~\ref{eq_pivo1}-\ref{eq_pivo3}, we neglected direct hydrodynamic interactions
between the two flagella and approximated the active driving forces by 
$Q_L(\varphi)=\omega_0\Gamma_{LL}(\varphi,\varphi)_{|\psi=0}$ and 
$Q_R(\varphi)=\omega_0\Gamma_{RR}(\varphi,\varphi)_{|\psi=0}$.
The coupling functions are defined as
$\ol{\mu}(\varphi)=-\Gamma_{L\psi}(\varphi,\varphi)/\Gamma_{LL}(\varphi,\varphi)_{|\psi=0}$,
$\ol{\nu}(\varphi)=-\Gamma_{\psi L}(\varphi,\varphi)_{|\psi=0}$, and
$\ol{\rho}(\varphi,\varphi)=\Gamma_{\psi\psi}(\varphi_L,\varphi_R)_{|\psi=0}$.
This choice retains the key nonlinearities of the full equation of motion, see also figure \ref{fig_gh}.
Equation~\ref{eq_pivo1} states that pivoting of the flagellar basal apparatus with $\dot{\psi}>0$ slows down the
effective stroke of the left flagellum (and speeds up the right flagellum).
For synchronized flagellar beating, there will be no pivoting of the flagellar base.
For asynchronous beating, however, the flagellar base will be rotated out of its
symmetric rest position by an angle $\psi$ if the stiffness $\kpivo$ is not too large.
Any pivoting motion of the flagellar base during the beat cycle
changes the hydrodynamic friction forces that oppose the flagellar beat,
which in turn can either slow down or speed up the respective flagellar beat cycles, 
and thus restore flagellar synchrony.

To gain further analytical insight, 
we study the response of the dynamical system \ref{eq_pivo1}-\ref{eq_pivo3}
after a small perturbation $0<\delta(0)\ll 1$.
To leading order in $\delta=\varphi_L-\varphi_R$, we find (with $\varphi\approx\omega_0 t$)
\begin{align}
\kpivo\psi+\rho(\varphi,\varphi)\dot{\psi} & = -d[\ol{\nu}(\varphi)\delta]/dt, \\
\dot{\delta} & = -2\ol{\mu}(\varphi)\dot{\psi}, 
\end{align}
In the biologically relevant case
of a relatively stiff basal anchorage of the flagellar basal apparatus with $\kpivo\gg\rho\omega_0$,
we find for the synchronization strength a result analogous to equation \ref{eq_lelastic}
\begin{equation}
\label{eq_lpivo}
\ol{\lambda} = - \oint_0^{2\pi}\!\! d\varphi\,\, \frac{\ol{\mu}(\varphi)\ol{\nu}''(\varphi)}{\kpivo/\omega_0}.
\end{equation}











\forcenewpage

\section{Supplementary Figures}

\forcenewpage

\forcenewpage

\forcenewpage

\forcenewpage



\forcenewpage

\forcenewpage

\begin{figure}
\includegraphics[width=8.5cm]{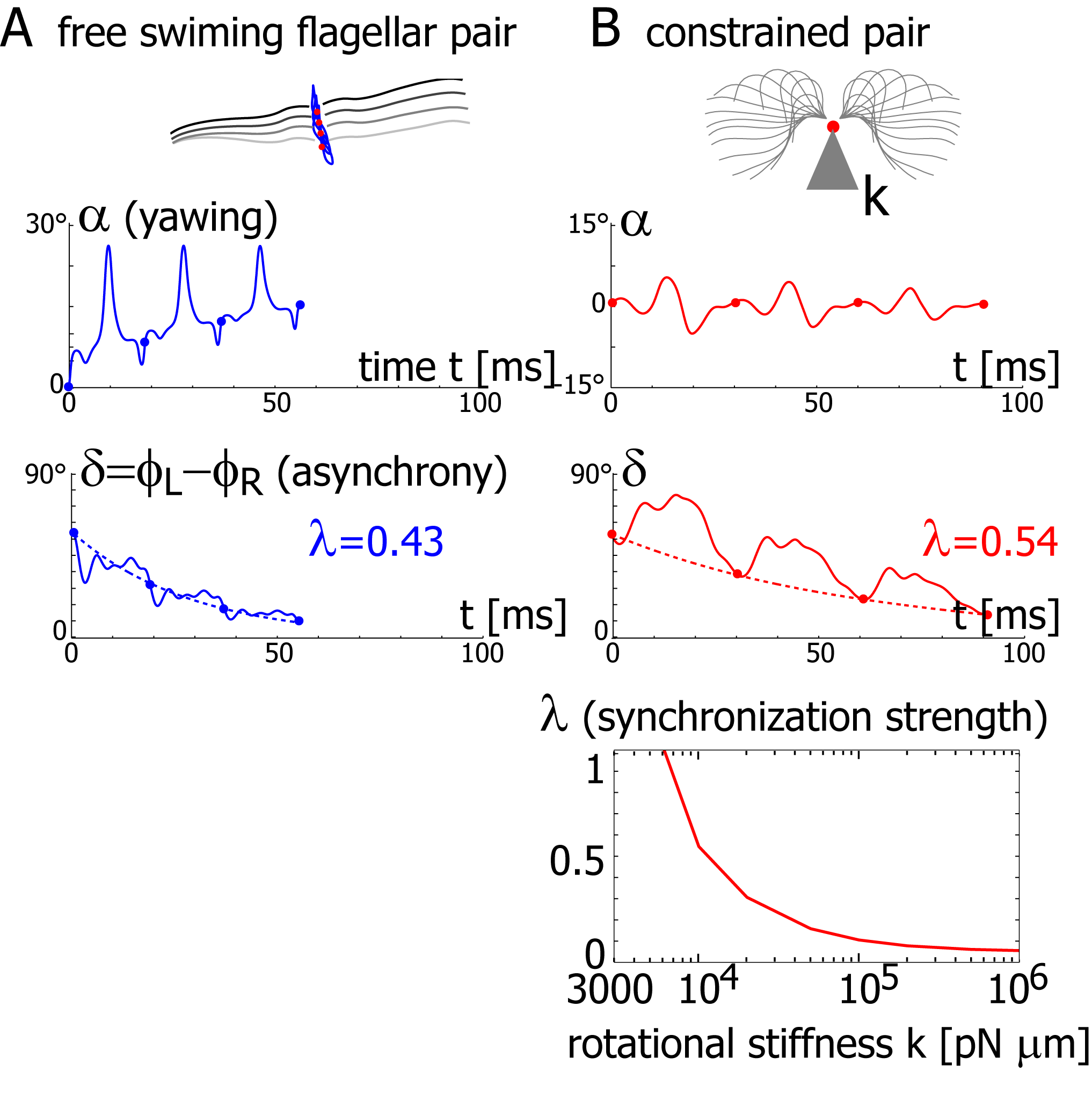}
\caption{
\label{fig_sync_isolated_pair}
Theory of flagellar synchronization for an isolated flagellar pair.
Inspired by experiments by Hyams \textit{et al.} \cite{Hyams:1975}
reporting synchronization in isolated flagellar pairs, we computed
the swimming and synchronization behavior of a flagellar pair with cell body removed.
For the computations, we used flagellar shapes and flagellar driving forces $Q_j(\varphi)$, $j=L,R$,
determined from an intact cell, see figure \ref{fig_dof}.
\textbf{A.}
For a free swimming flagellar pair,
we observe a characteristic yawing motion of the flagellar pair characterized by $\alpha(t)$,
if the two flagella are initially out of synchrony.
The flagellar phase difference $\delta$ is found to decrease with time (solid line),
approximately following an exponential decay (dotted line).
This implies that the in-phase synchronized state is stable with respect to perturbations.
Each completion of a full beat cycle of the left flagellum is marked by a dot.
\textbf{B.}
To mimic experiments where external forces constrain the motion of the flagellar pair,
we simulated the idealized case of a pair that cannot translate,
while yawing of the pair is constricted by an elastic restoring torque $Q_\alpha=-\kyaw\alpha$
that acts at the basal apparatus (red dot).
As in the case of a free-swimming pair, 
the flagellar phase difference $\delta$ decays with time, indicating stable synchronization.
In the case of a constrained cell, the synchronization strength $\lambda$ strongly depends
on the clamping stiffness $\kyaw$.
Parameters: $2\pi/\omega_0=30\,\mathrm{ms}$, $\kyaw=10^4\,\mathrm{pN}\,\mu\mathrm{m/rad}$.
To enhance numerical stability, we added a small constant
$\kappa=10\,\mathrm{pN}\,\mu\mathrm{m}\,\mathrm{ms}$
to the flagellar friction coefficients,
$\Gamma_{jj}(\varphi_L,\varphi_R)$, $j=L,R$,
which corresponds to internal dissipation \cite{Friedrich:2012c}.
}
\end{figure}

\forcenewpage

\forcenewpage

\end{document}